\documentclass[a4paper,final,5p,times,twocolumn]{elsarticle}
\usepackage[T1]{fontenc}
\usepackage[latin9]{inputenc}
\usepackage{amsthm}
\usepackage{amsmath}
\usepackage{amssymb}
\usepackage{graphicx}

\def\simleq{\mathrel{\smash{\mathop{\raise2pt\hbox{$<$}}\limits_%
   {\smash{\raise4pt\hbox{$\sim$}}}}\vphantom\leq}}
\def\simgeq{\mathrel{\smash{\mathop{\raise2pt\hbox{$>$}}\limits_%
   {\smash{\raise4pt\hbox{$\sim$}}}}\vphantom\geq}}

\pdfpageheight\paperheight
\pdfpagewidth\paperwidth

\providecommand{\tabularnewline}{\\}

\journal{Computer Physics Communications}

\begin{document}

\begin{frontmatter}




\title{Towards optimal explicit time-stepping schemes for the gyrokinetic
equations}

\author{H.~Doerk\corref{hkd} and F.~Jenko}
\ead{Hauke.Doerk@ipp.mpg.de}
%\author{F.~Jenko\corref{fsj}}
%\ead{Frank.Jenko@ipp.mpg.de}
\cortext[hkd]{corresponding author}
\address{Max-Planck-Institut für Plasmaphysik, Boltzmannstraße
2, D-85748 Garching, Germany}

\begin{abstract}
The nonlinear gyrokinetic equations describe plasma turbulence in
laboratory and astrophysical plasmas.
%old
%To solve these equations, massively parallel codes are developed and run on present-day supercomputers.
%The goal of this paper is to improve the efficiency of such computations,
%thereby allowing for physically more comprehensive studies. 
%begin referee suggestion
To solve these equations, massively parallel codes have
been developed and run on present-day supercomputers. This paper describes
measures to improve the efficiency of such computations, thereby making
them more realistic.
%end referee suggestion
Explicit Runge-Kutta schemes are
considered to be well suited for time-stepping. Although the numerical
algorithms are often highly optimized, performance can still be improved
by a suitable choice of the time-stepping scheme, based on spectral
analysis of the underlying operator. Here, an operator splitting technique
is introduced to combine first-order Runge-Kutta-Chebychev schemes
for the collision term with fourth-order schemes for the remaining
terms. In the nonlinear regime, based on the observation of eigenvalue
shifts due to the (generalized) $E\times B$ advection term, an accurate
and robust estimate for the nonlinear timestep is developed. The presented
techniques can reduce simulation times by factors of up to three in
realistic cases. This substantial speedup encourages the use of similar
timestep optimized explicit schemes not only for the gyrokinetic equation,
but also for other applications with comparable properties.
\end{abstract}
%\maketitle

\end{frontmatter}

\section{Introduction}

Gyrokinetic simulation codes are a common tool for obtaining\emph{
ab-initio} predictions of turbulence properties in strongly magnetized
high-temperature plasmas.\citep{GarbetIdomura2010,Krommes2012} Such
plasmas are present in magnetic confinement fusion devices, and in
astrophysics.
%Many phenomena such as the cross-field transport of heat, particles, and momentum are presently understood as the effect of plasma turbulence.
Gyrokinetic theory describes the time evolution
of each species' particle distribution function $f$ in five-dimensional
phase space (one velocity space variable, the gyro-angle, is averaged
out). Obtaining a solution of this nonlinear partial integro-differential
equation generally requires high-performance computing. In the past
decades, gyrokinetic codes have become substantially more realistic by
applying higher numerical resolution and by moving to more comprehensive
physics models. For example, the effect of collisions is formally weak in dilute
high-temperature plasmas and thus has often been neglected. Today,
one realizes that including a suitable collision operator in gyrokinetic
turbulence is not only required for a physically correct entropy balance,\citep{Sugama2009}
but can also greatly influence the turbulence level--through damping
of zonal flows--or even change the turbulence regime by modifying the
growth rate of certain types of microinstabilities.\citep{Ernst2004,Applegate2007,Xiao2007,Guttenfelder2012,HatchSMT2013}
Since more realistic physics models require increased computational effort,
progress is enabled by the availability of more powerful computers
and by the use of advanced algorithms, the importance of the latter often being underestimated.

Three classes of gyrokinetic turbulence codes (particle-in-cell, semi-Lagrangian, and Eulerian) exist. 
Here, the Eulerian approach, which became popular approximately fifteen years ago, is considered.
Several major code projects exist in this area, for instance GENE \citep{JenkoCoPhC2000,Jenko00ETGpop,Dannert05PoP,Goerler11gene},
GS2 \citep{Kotschenreuther1995Gs2,Dorland2000Gs2}, GYRO \citep{CandyWaltz2003JCoPhGyro,CandyWaltz2003PRLGyro},
GKW \citep{Peeters2009GKW}, and AstroGK \cite{Numata2010}. The common basic procedure is the so-called method of lines: 
After discretizing phase space on a fixed grid, the resulting large system of 
ordinary differential equations is evolved with a time integration scheme.
However, the choice of algorithms can differ substantially.
Besides various possible choices for phase space grids and the representation of derivatives on those grids,
time discretization is performed in several ways, see Ref.~\citep{Maeyama2013CoPhC} for a useful overview.
Operator splitting techniques for the collisional term are used in GYRO, GS2 and AstroGK. Some codes (like GS2)
even choose to split off the nonlinear term from linear dynamics,
while others avoid splitting to treat these terms on an equal level.
Moreover, implicit, as well as explicit schemes are applied. While GS2 (and AstroGK) treat all linear terms implicitly,
the GYRO algorithm splits off fast linear terms (the parallel electron dynamics) in an implicit-explicit (IMEX) fashion. 
Here, we focus on fully explicit time integration, as employed in GENE and GKW, for example.
Explicit methods offer the advantages of an excellent performance on massively parallel systems 
and the straightforward implementation of nonlinear terms.
The drawback is a strict stability limit that is set on the timestep $\Delta t$, which depends on the fastest dynamics in the system.
A major advance from gyrokinetic theory is to analytically remove extremely fast
timescales like compressional Alfvén waves or particle gyromotion, leaving only relevant dynamics and enabling an explicit treatment.
One of the fastest remaining terms is then given by the (generalized) nonlinear drift velocity $v_{\chi}=E_{\chi}\times B$
that combines electric and magnetic field fluctuations.
When this nonlinear advection limits the timestep according to a Courant-Friedrichs-Lewy (CFL) relation $\Delta t\simleq\Delta x/v_{\chi}$,
%for the spatio-temporal resolution of the numerical treatment of (advective) partial differential equations,
\citep{CourantFriedrichsLewy1928} fully explicit schemes are likely to be the more efficient choice
(particularly in view of increasing problem size)%, the advantages of explicit schemes are striking.
.\citep{JenkoCoPhC2000}

It is sometimes stated that collisions require an implicit treatment, since the explicit diffusive timestep limit would be too strict.\citep{Barnes2009}
However, we find severe restrictions only for rather large collision frequencies (in the tokamak edge, for example) or for very high velocity resolution.
In this work, we introduce a splitting scheme involving Runge-Kutta-Chebychev (RKC) schemes with extended real stability boundary,\citep{HouwenSommeijer1980,Verwer1996} 
which enables an explicit treatment of a sophisticated collision operator even in these extreme cases.
Partitioned RKC schemes have recently been developed which are also stable for advective terms, involving, however, a larger number of operator evaluations per step.\citep{ZbindenSIAM2011,AbdulleVilmart2013JCoPh}

In principle, accuracy limits can also be imposed on the timestep. In this context, we note that the overall numerical accuracy of gyrokinetic simulations is generally strongly restricted by the grid resolution in five-dimensional phase space. A relative error tolerance of approximately $10^{-3}$ is already considered to be sufficient, even for linear simulations. Nonlinear simulations are subject to statistical errors of the order of 10\%, underlining the fact that long simulation times rather than highly accurate steps are needed.
In consequence, the use of low-order time integration schemes is well justified to speed up computations.

In this paper, a detailed analysis of the spectral properties of the discretized system allows us to identify a
class of highly efficient first-order explicit schemes (with largely extended stability boundaries),
which we apply to the gyrokinetic code GENE.
%substatial speedup is aimed at in three ways.
%First, schemes of extended stability limit are applied, where necessary.
%Second, the number of operator evaluations per step is reduced to a minimum.
%Third, the timestep estimate in nonlinear simulations is improved to approach the exact limit as close as possible.
%
%top of that, explicit RK schemes have the property that the stability
%boundary can be shaped and extended by adding few additional evaluations
%of the operator. This naturally suggests the development of timestep
%optimized RK schemes that are tailored to the requirements of the
%considered system, as long as the enhanced efficiency does not effect
%the computation accuracy (see Refs.~\citep{Verwer1996,Mead1999},
%for example).
%In this paper, we discuss possibilities for maximizing the timestep
%based on the spectral properties of the linear and the nonlinear gyrokinetic
%operator and apply them to the turbulence code \textscGENE.\citep{Jenko00ETGpop,Dannert05PoP,Goerler11gene}
The remainder of this paper is organized as follows. The relevant equations are summarized in Section~\ref{sec:gyrokinetics}
and timestep limiting physics is discussed.
In Section~\ref{sec:Stability-of-explicitRK}
we introduce relevant explicit RK schemes and review their stability conditions.
%conditions. In Section~\ref{sec:Operatorsplit}, splitting techniques
%are discussed, which allow to apply well-suited time-stepping schemes
%to the individual parts of the operator.
In Section~\ref{sec:Operatorsplit}, the efficiency and accuracy of splitting techniques
are discussed, which allow time-stepping schemes to be tailored
to the individual parts of the operator.
Finally, in Section~\ref{sec:nlev} we address the timestep restrictions in nonlinear simulations. We show that the $E_{\chi}\times B$ advection shifts the eigenvalues along the imaginary axis, which is relevant for the stability limit.  This observation forms the basis of a greatly improved estimate of the nonlinear timestep.
Overall, these two methods of (i) operator splitting and (ii) an
improved timestep estimate enhance the code efficiency by up to a
factor of three in realistic cases. 
%In view of an already highly optimized code, this speedup is significant.
Since the code was already highly optimized, this speedup is significant.

\section{The gyrokinetic equations \label{sec:gyrokinetics}}

The gyrokinetic equation
\begin{align}
\partial_{t}g=G[t,g]=N[\bar{\chi},g]+L[g]+C[g]\label{eq:GK_symbolic}
\end{align}
describes the time evolution of the (modified) perturbed gyrocenter
distribution $g$ for each plasma species in $\{x,y,z,v_{\|},\mu\}$
phase space. The notation
\begin{align*}
\bar{\chi} & =\bar{\phi}_{1}-v_{\|}\bar{A}_{1\|}+\mu\bar{B}_{1\|}\;\qquad
f=g+v_{\|}\bar{A}_{1\|}F_{0}\\
F_{0} & =\frac{n_{0}}{\pi^{3/2}v_{T}^{3}}\exp\left[\left(v_{\|}^{2}+\mu B_{0}\right)/T_{0}\right].
\end{align*}
introduces the fluctuating potential $\bar{\chi}$, consisting of
electrostatic perturbations $\bar{\phi}_{1}$ and magnetic perturbations
$\bar{A}_{1\|}$ and $\bar{B}_{1\|}$, where the overbar denotes
a gyroaverage. The gyrocenter distribution is split into a background
(Maxwellian) distribution $F_{0}$ and a small fluctuating part $f$. The
background magnetic field is $B_{0}$ and the background density $n_{0}$,
temperature $T_{0}$, thermal velocity $v_{T}=\left(2T_{0}/m\right)^{1/2}$
and particle mass $m$ are given for each plasma species. The gyrokinetic
version of Maxwell's equations is used to compute a self-consistent
fluctuating potential from $g$, which closes the system of equations.
We refer to Refs.~\citep{Krommes2012,Jenko00ETGpop,Goerler11gene}
for a detailed description and derivation.

Eq.~\eqref{eq:GK_symbolic} is symbolically written as the sum of
three integro-differential operators whose physical meaning is briefly
discussed in the following. The linear terms $L[g]$ contain parallel
advection along the magnetic field lines, as well as perpendicular
drifts such as curvature and $\nabla B$ drifts, and temperature and
density gradient terms.
%Unless periodic boundary conditions are used in the radial direction $x$, sources and sinks are included in $L$ as well.
The nonlinear term $N[\bar{\chi},g]$ describes turbulent re-distribution of free energy
due to perpendicular $E_{\chi}\times B$ advection, where the generalized fluctuating field is defined as $E_{\chi}=-\nabla\bar{\chi}$.
Finally, the linearized Landau-Boltzmann
collision operator $C[g]$ describes diffusion and dynamical friction
in velocity space, including back-reaction terms that ensure conservation
of particles, momentum, and energy. Details of the implementation
of the collision operator in GENE can be found in Refs.~\citep{diss_flm,diss_hkd2013}.
%We note that in the considered model, the collision operator requires
%the gyrocenter distribution $f$ as an input. Also some terms of the
%linear operator require $f$ or $\bar{\chi}$ directly, so that solving
%for $f$ and $\bar{\chi}$ from $g$ is necessary at every timestep,
%even if the nonlinear term is not included. 

For numerical solution, Eq.~\eqref{eq:GK_symbolic}
is discretized on a fixed grid in phase space, where common techniques
from computational fluid dynamics, such as spectral methods, finite
differencing, finite element, and finite volume schemes can be used.
This results in a large system of ordinary differential equations
for the time evolution of the state vector $g$. When non-dissipative
differencing schemes are employed, as is the case with the GENE
code, it may be necessary to add hyperdiffusion terms to $L[g]$ that
remove unphysical grid-size oscillations in some phase space directions.\citep{Pueschel2010,Maeyama2013}

One way of solving this space-discretized system is to perform initial value computations,
for which we consider Runge-Kutta (RK) schemes here. 
In the nonlinear case, we desire to find a statistically stationary turbulent state.
Linear initial value computations yield the fastest growing solution (sometimes referred to as a mode),
which constitute the driving force for plasma turbulence and are thus of great interest.
Typical growth rates and frequencies are of the order of $c_{s}/L_{\mathrm{ref}}$,
where $c_{s}=(T_{e}/m_{i})^{1/2}$ denotes the ion sound speed and
$L_{\mathrm{ref}}$ is a typical macroscopic scale length, often set to the tokamak major radius.
Additionally, the linearized system can be formulated as an eigenvalue problem.
In this context, GENE features the use of optimized iterative algorithms provided
by the SLEPc package,\citep{Hernandez2005,SLEPChomepage,Kammerer2008,Roman2010slepc,Merz2012}
which select a subset of eigenvector-eigenvalue pairs
$\{g_{i},\lambda_{i}\}$ that fulfill some user-specified criteria.
For convenience, we split the complex eigenvalue $\lambda=\gamma+i\omega$
into a growth rate $\gamma$ and a frequency $\omega$. 
The eigenvalues of largest magnitude $|\lambda_{i}|$ are quickly found
(for example by Krylov-Schur subspace iteration),
which proves extremely useful for the exact computation of the maximum stable timestep for initial value simulations.
Due to the shape of the spectrum, obtaining the fastest growing solution with SLEPc is more cumbersome,
but can still be faster than a corresponding initial value simulation.
Moreover, subdominant and marginally stable solutions only become accessible by such eigenvalue computations.
Finally, GENE can also compute the full spectrum (using ScaLAPACK routines),
but this is only feasible for small problems.

%Importantly, only a few exponentially growing ($\gamma>0$) physical solutions
%exist, which have a moderate characteristic frequency $\omega\sim c_{s}/L_{\mathrm{ref}}$.
%Here, $c_{s}=(T_{e}/m_{i})^{1/2}$ denotes the ion sound speed and
%$L_{\mathrm{ref}}$ is a typical macroscopic scale length, often set
%to the tokamak major radius in fusion applications. These unstable
%waves constitute the driving force for plasma turbulence and are thus
%of greatest interest. However, the timestep is usually not limited
%by accuracy constraints, but rather by the stability of other parts
%of the spectrum with $\gamma\leq0$. The explicit timestep must be
%small enough to ensure that these solutions are not numerically destabilized.
%
As we will see in Sec.~\ref{sec:Stability-of-explicitRK},
the maximum stable timestep for Runge-Kutta methods is determined by the spectral properties of the underlying operator. 
Focussing on the linear case first, either the fastest oscillating $\omega_{\mathrm{max}}$
or the most damped $\gamma_{\mathrm{min}}=$min{[}$\mathrm{Re}(\lambda)${]}
solutions are typically most restrictive.
Let us briefly summarize physical mechanisms behind these extreme
eigenvalues. 
Importantly, the integro-differential character of the parallel advection term
$\partial_t g=v\partial_z g + v\partial_z \chi$ does not allow for a rigorous CFL approach
of the form $\lambda_{\mathrm{max}}=k_{\mathrm{max}}v$, since $\chi$ is computed from $g$ integrals.
Here, $v$ is an advection velocity and $k_{\mathrm{max}}$ is the largest wavenumber in the system.
%, which could be
%given by the Nyquist number $\pi/\Delta$ for constant grid-spacing $\Delta$ and a domain size of $2\pi$, for instance.
A popular example for the origin of very high-frequency
(and timestep limiting) solutions are kinetic shear Alfvén waves.
In simplified slab geometry (and in the relevant low-$\beta_e$ limit),
the dispersion relation reads
\begin{align}
\omega_{\mathrm{KSA}}^{2}=\frac{1}{\mu_{e}(k_{\perp}\rho_{s})^{2}+\beta_{e}}k_{\parallel}^{2}c_{s}^{2}\,,\label{eq:KSAW}
\end{align}
where $k_{\perp}$ is a perpendicular wavenumber and $k_{\parallel}$
is a parallel wavenumber.\citep{WWLeePoP2001,Dannert2004} Here, the
electron to ion mass ratio $\mu_{e}=m_{e}/m_{i}$, the electron beta
$\beta_{e}=4\pi n_{e0}T_{e0}/B_{0}^{2}$, the ion sound gyroradius
$\rho_{s}=c_{s}/\Omega_{i}$ and the ion cyclotron frequency $\Omega_{i}=(eB_{0})/(m_{i}c)$
are introduced. The $\beta_{e}$ parameter controls the response in
Amp\`ere's law, whereas electrostatic models use $\beta_{e}=0$. We
observe that as \textbf{$\beta_{e}$} approaches zero, the frequency
$\omega_{\mathrm{KSA}}\sim k_{\|}/(k_{\perp}\rho_{s})$ can become
very large. Indeed, setting $k_{\|\mathrm{max}}L_{\mathrm{ref}}\sim\pi/\Delta z\sim10$,
$k_{\perp}\rho_{s}\sim0.05$, $\mu_{e}=1/3600$ and $\beta_{e}=0$
we obtain $\omega_{\mathrm{KSA}}\sim4000\,(c_{s}/L_{\mathrm{ref}})$, about four
orders of magnitude larger than the typical values for growth or damping
rates. Fortunately, even small values of $\beta_{e}$ prevent the
divergence of $\omega_{{\rm KSA}}$, so that it can be beneficial
to include electromagnetic effects for kinetic electron simulations,
even if the dominant physics is of electrostatic nature. In the opposite
limit of $\beta_{e}/\mu_{e}\simgeq(k_{\perp}\rho_{s})^{2}$, which
is more relevant to actual fusion plasmas, Eq.~\eqref{eq:KSAW} transitions
into the classical Alfvén wave dispersion relation $\omega_{{\rm A}}^{2}=v_{{\rm A}}^{2}k_{\|}^{2}$
with $v_{{\rm A}}=c_{s}/\beta_{e}^{1/2}$ denoting the Alfvén velocity.
Also the parallel streaming of electrons is often
relevant, even if field-aligned coordinates are used. The characteristic
frequency is given as $\omega_{\|}\sim k_{\|}v_{\parallel e}$,
which can be linked to a CFL condition.
In typical fusion experiments, the electron thermal velocity is larger
or comparable to the Alfvén velocity ($v_{\|e}/v_{{\rm A}}\sim v_{te}/v_{{\rm A}}\sim(\beta_{e}/\mu_{e})^{1/2}\simgeq1$),
so that the CFL condition for kinetic electrons is usually more restrictive
than the limit due to Alfvén waves.

A third notable source of high frequency solutions is linked to magnetic
curvature and $\nabla B$ drifts, which are (roughly) proportional
to particle energy $\epsilon=mv^{2}/2$ and perpendicular wavenumber
$k_{\perp}$. Thus, if either highly energetic particles or very high
wavenumbers are involved, these drifts are expected to play a relevant role.
%
%%%---------------------figure 1 -------------------------%%%
\begin{figure}
\begin{centering}
(a)\hspace{0.5\columnwidth}(b)
\par\end{centering}

\includegraphics[width=0.45\columnwidth]{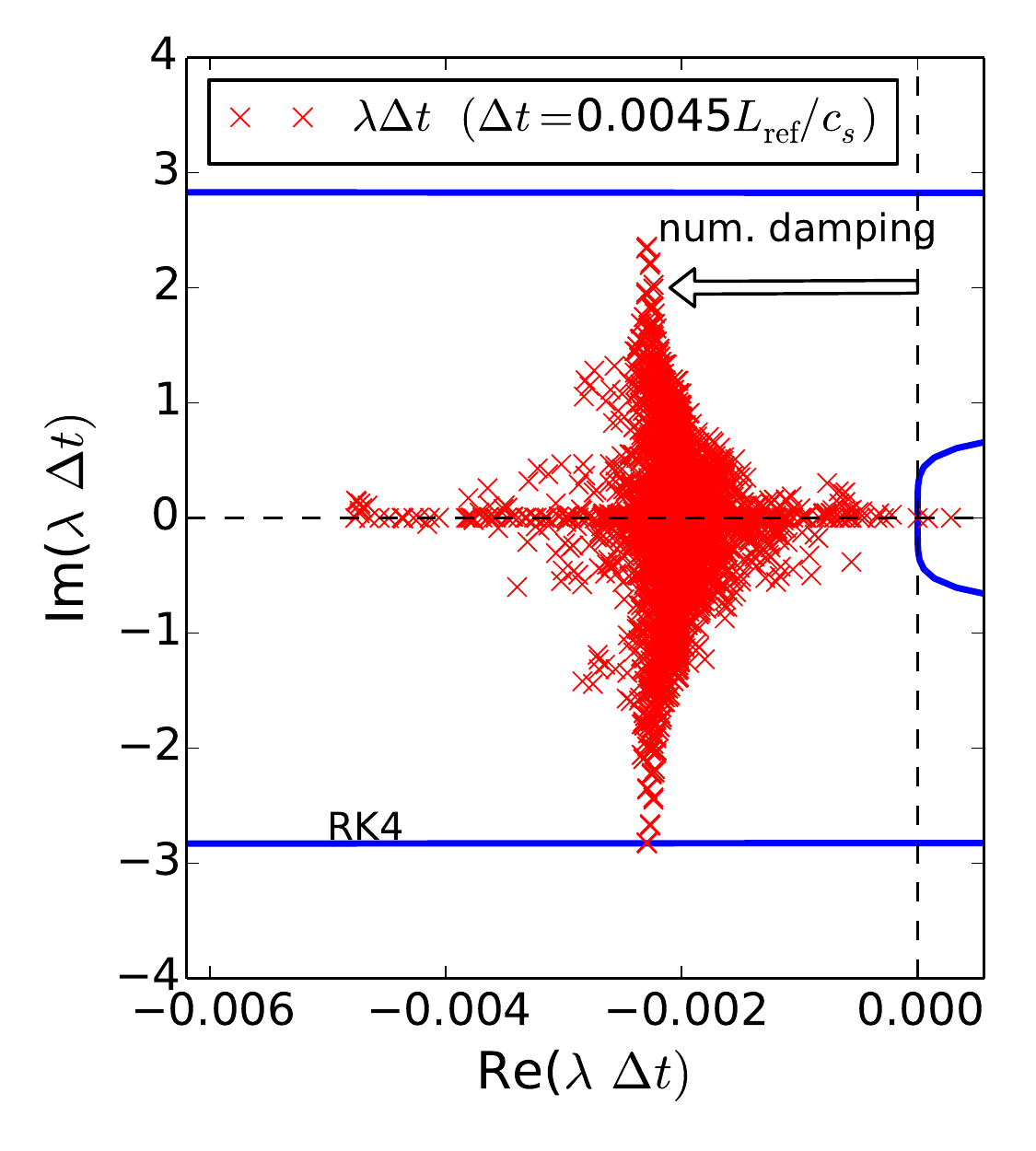}\hfill{}\includegraphics[width=0.45\columnwidth]{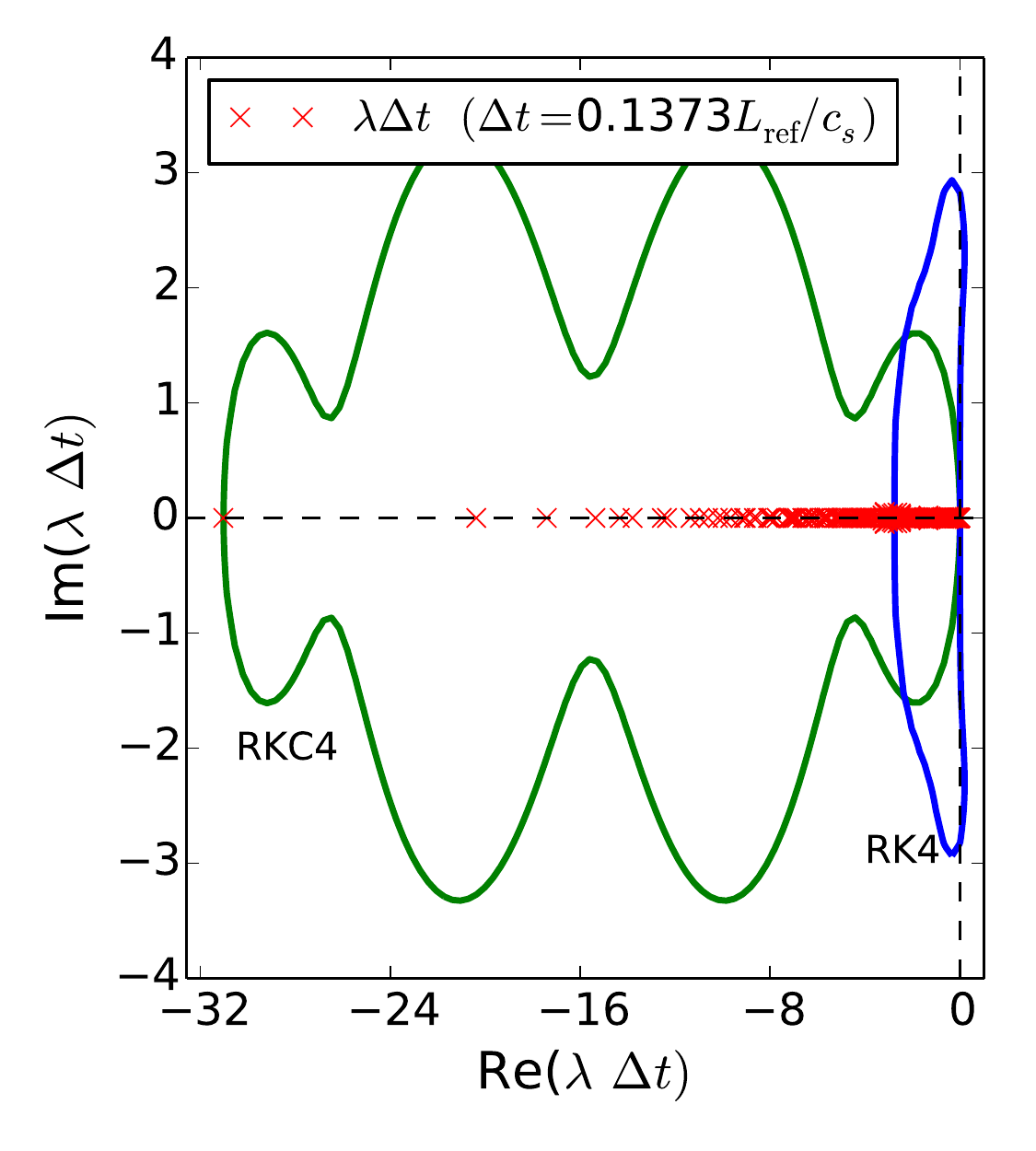}
%nlev/evclouds/allev2

\begin{centering}
(c)\hspace{0.5\columnwidth}(d)
\par\end{centering}

\includegraphics[width=0.45\columnwidth]{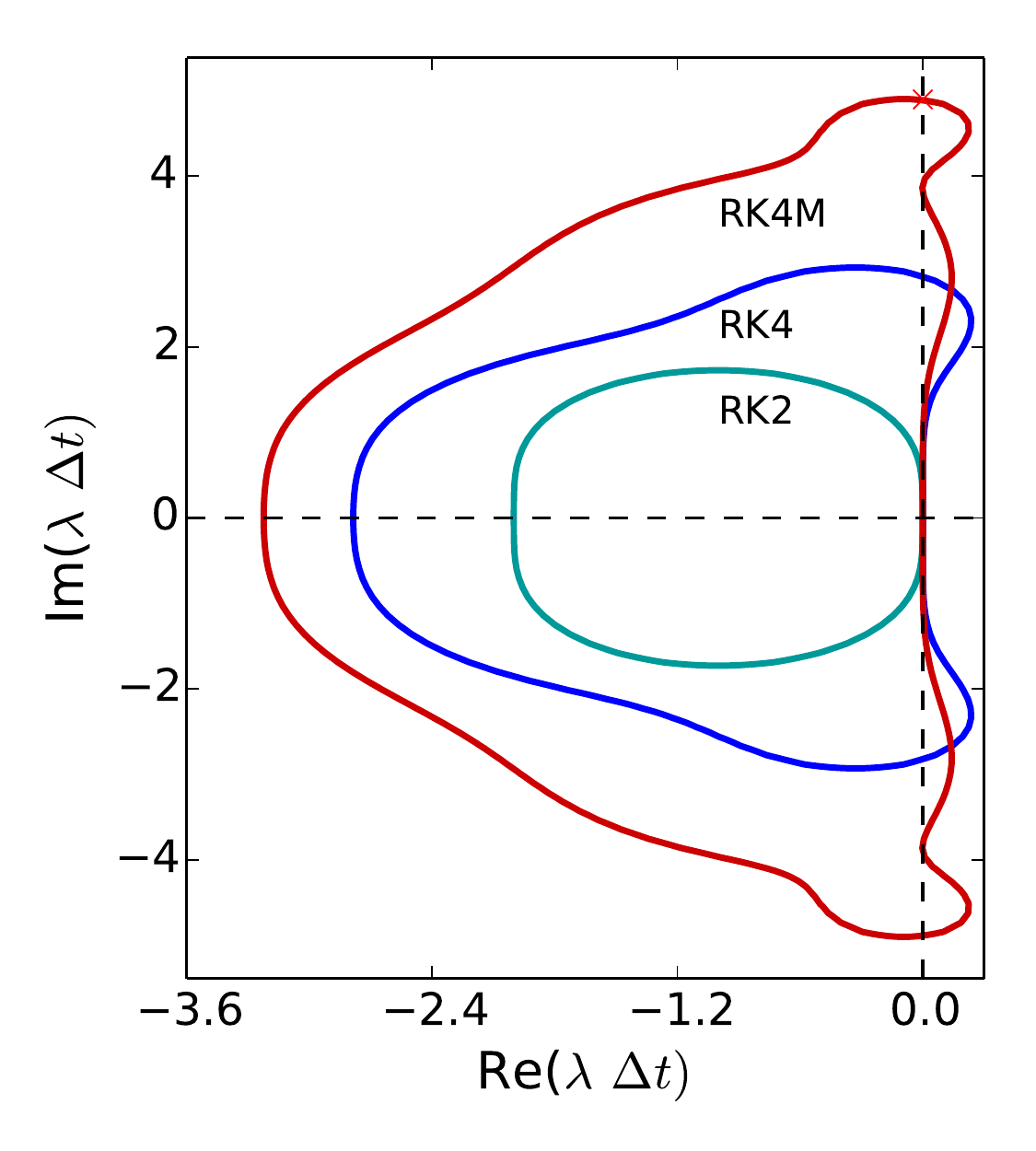}\hfill{}\includegraphics[width=0.45\columnwidth]{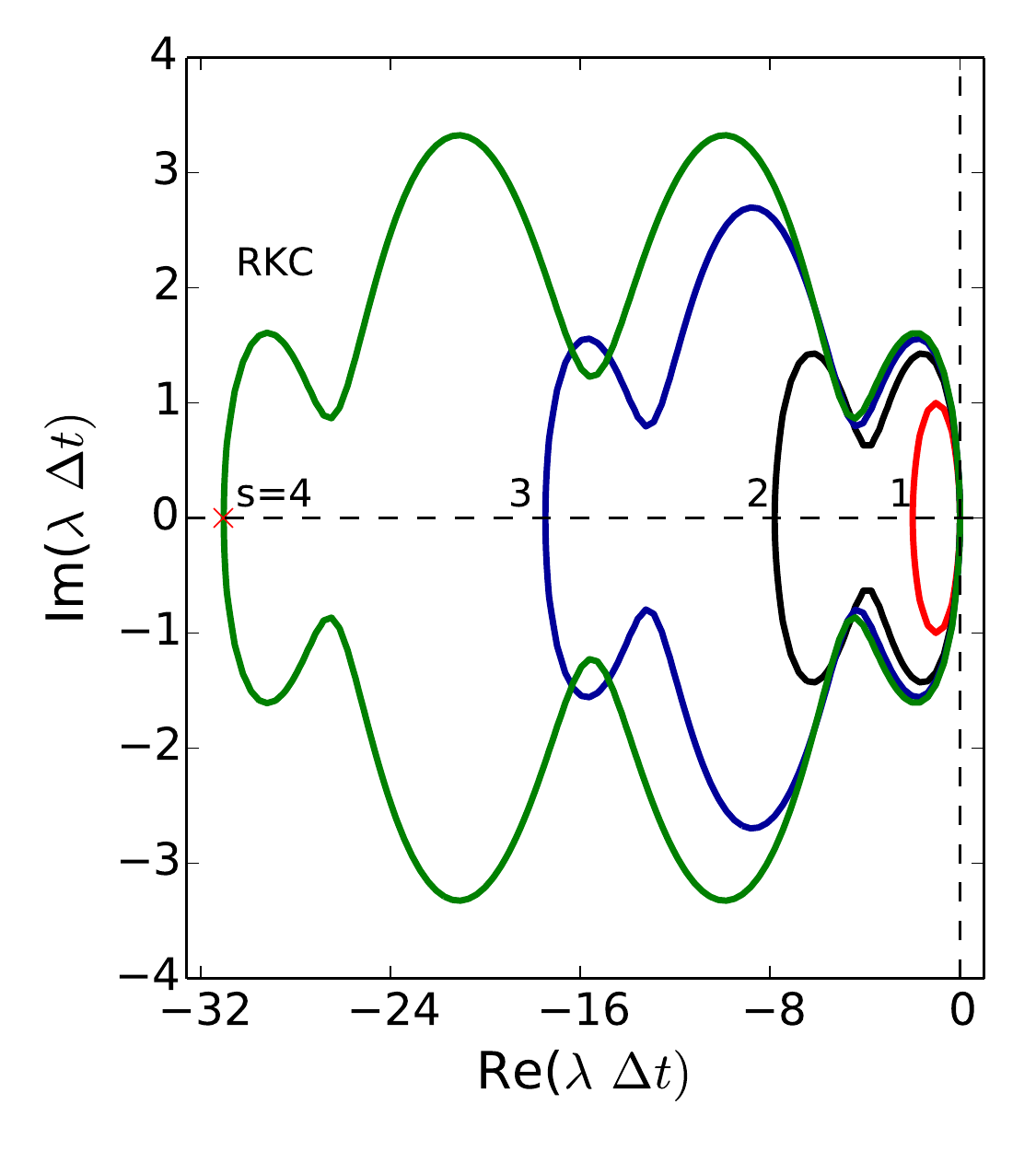}

\caption{(a) Spectrum of the collisionless, linearized gyrokinetic operator $L$, scaled with
$\Delta t_{{\rm RK4}}$ and (parts of) the stability boundary of the
RK4 scheme. Note that $\mathrm{Im}(\lambda)\gg Re(\lambda)$.
(b) Spectrum of the collision operator,
scaled with $\Delta t_{{\rm RKC4}}$ and stability boundaries of the
RK4 and RKC4 schemes (for $\Delta t=\Delta t_{{\rm RKC4}}$).
Here, the eigenvalues have a dominating real part, where the RKC4
stability region has a much larger extent. (c) Stability boundaries
of the RK2, the RK4 and the RK4M schemes. (d) First-order RKC stability boundaries
up to four stages.\label{fig:RK4_RKC_stability}}
\end{figure}

Of course, the above considerations are based on simplified versions
of the gyrokinetic equation, or even on single terms. In the general,
more comprehensive case, all these terms are coupled and one has to numerically compute the spectrum.
The result (using ScaLAPACK) is shown in Fig.~\ref{fig:RK4_RKC_stability}(a) for
the discretized, collisionless linear gyrokinetic operator.
Indeed, the eigenvalue spectrum is stretched along the imaginary axis.
%which also reflects the fact that non-dissipative discretization is chosen.
One generally finds the highest frequency either at smallest or largest wavenumber $k_{y}$,
which is consistent with the phenomena discussed above. In the appropriate
limits, the magnitude of the numerically computed frequency scales
as expected.
While hyperdiffusion terms can be necessary to stabilize spurious grid-size
oscillations, their effect on the eigenvalue spectrum is just a comparably small shift along the 
negative real axis, which has no big impact on stability considerations.
It is important to note that upwind (or other dissipative) discretization methods can strongly distort the spectrum,
and that dissipation on low $k_{y}$ potential fluctuations $\bar{\phi}$ can cause 
large negative eigenvalues and should thus be avoided.\citep{Dannert2004}

%Thus far, we have discussed wave phenomena with a certain frequency
%$\omega$. Often, these waves are exponentially damped with a rate
%$\gamma<0$, which in our definition is the real part of the eigenvalue.
%Only few solutions are driven unstable by physical mechanisms with
%a typical growth rate of $0<\gamma\simleq1$. These are not of concern
%for the numerical stability of explicit time-stepping schemes. 
%Damping is introduced by numerical dissipation (e.g. hyperdiffusion/hypercollision
%terms or some discretization schemes) and the physical collision operator.
%While hyperdiffusion terms are needed to stabilize spurious grid-size
%oscillations of centered finite differencing schemes, their effect
%on the eigenvalue spectrum is just a comparably small shift along the 
%negative real axis. 
%Nevertheless, applying dissipation on
%low $k_{y}$ potential fluctuations $\bar{\phi}$ should be avoided,
%because this can lead to single strongly damped solutions that require
%very small explicit timesteps.\citep{Dannert2004} 
In the flux-tube limit, collisional diffusion (a sink of fluctuation entropy) provides
the only physically motivated damping in the gyrokinetic system.\citep{Sugama1996entropy}
The numerically computed spectrum of the collision operator is shown
in Fig.~\ref{fig:RK4_RKC_stability}(b); the eigenvalues are distributed
along the negative real axis.%
\footnote{Imaginary parts that arise from discretization and boundary conditions
are negligible.}%
The maximum damping rate is proportional to the collision frequency
$\nu_{{\rm c}}$ and roughly proportional to $\Delta v_{\|}^{-2}$,
which reflects the diffusive character of this term. Since the velocity-dependent
collision rate diverges for $v=0$, it is essential to distribute
$v_{\|}$ grid points symmetrically around $v_{\|}=0$, with no grid
point at the origin. Even if this is done, collisions can dominate
the timestep, particularly towards the edge of tokamak devices (like
{ASDEX} Upgrade~\citep{AUG2011NucFu}), where the collisionality is larger
compared to core plasmas, and stronger flux-surface shaping can require
a finer velocity grid.

%The full gyrokinetic equation Eq.~\eqref{eq:GK_symbolic} can not
%be written as an eigenvalue problem because in the nonlinear term
Finally, including the nonlinear $E_{\chi}\times B$ term
\begin{align}
\label{eq:nonlinearity}
N[\bar{\chi},g] & = \frac{c}{\mathcal{C}_{xy}}\frac{B_{0}}{B_{0\|}^{*}}\left(\partial_{y}\bar{\chi}\partial_{x}g-\partial_{x}\bar{\chi}\partial_{y}g\right)\\
 & \equiv v_{\chi}^{x}\partial_{x}g+v_{\chi}^{y}\partial_{y}g\nonumber
\end{align}
prohibits a direct eigenvalue computation, since the advection velocities
$v_{\chi}^{x}$ and $v_{\chi}^{y}$ are computed self-consistently from $g$. 
In such cases, a common technique is to freeze $v_{\chi}$ at time $t_n$ (and maximize it in space) for computing a timestep estimate.
Details on this procedure and the combination of nonlinear and linear stability limits are given in Section~\ref{sec:nlev}.
Due to the advective character of $N[\bar{\chi},g]$, a frequency shift along the imaginary axis is found.
%Here, $\mathcal{C}_{xy}$ is a geometric factor of the field-aligned coordinate system.
%Nevertheless, the total stability limit will be determined by some combination of linear
%and nonlinear terms. 

\section{Stability properties of relevant explicit Runge-Kutta methods\label{sec:Stability-of-explicitRK}}
%
%----------------table 1--------------------%
\begin{table*}
\begin{centering}
{\footnotesize }%
\begin{tabular}{ccccccc}
\hline 
 & {\footnotesize $a_{1}$} & {\footnotesize $a_{2}$} & {\footnotesize $a_{3}$} & {\footnotesize $a_{4}$} & {\footnotesize $a_{5}$} & {\footnotesize $a_{6}$}\tabularnewline
\hline 
\hline 
{\footnotesize RK2} & {\footnotesize $0$} & {\footnotesize $0.8$} &  &  &  & \tabularnewline
{\footnotesize RK4} & {\footnotesize $0$} & {\footnotesize $0.5$} & {\footnotesize $0.5$} & {\footnotesize $1.0$} &  & \tabularnewline
{\footnotesize RK4M} & {\footnotesize $0$} & {\footnotesize $0.16791847$} & {\footnotesize $0.4829844$} & {\footnotesize $0.7054607$} & {\footnotesize $0.0929587$} & {\footnotesize $0.7621008$}\tabularnewline
\hline 
 & {\footnotesize $b_{1}$} & {\footnotesize $b_{2}$} & {\footnotesize $b_{3}$} & {\footnotesize $b_{4}$} & {\footnotesize $b_{5}$} & {\footnotesize $b_{6}$}\tabularnewline
\hline 
\textsc{\footnotesize RK2} & \textsc{\footnotesize $0.375$} & \textsc{\footnotesize $0.625$} &  &  &  & \tabularnewline
\textsc{\footnotesize RK4} & \textsc{\footnotesize $1/6$} & \textsc{\footnotesize{} $1/3$} & \textsc{\footnotesize $1/3$} & \textsc{\footnotesize $1/6$} &  & \tabularnewline
\textsc{\footnotesize RK4M} & \textsc{\footnotesize $-0.15108371$} & \textsc{\footnotesize $0.7538468$} & \textsc{\footnotesize $-0.3601660$} & \textsc{\footnotesize $0.5269677$} & \textsc{\footnotesize $0.0$} & \textsc{\footnotesize $0.2304351$ }\tabularnewline
\hline 
\end{tabular}
\par\end{centering}{\footnotesize \par}

\caption{Coefficients for a two stage RK2, a four stage RK4 and a six stage
RK4M method.\label{tab:RK4_table}}
\end{table*}
%----------------table 1--------------------%
%
%The gyrokinetic equation $\partial_{t}g=G[t,g]$, Eq.~\eqref{eq:GK_symbolic},
%involves the nonlinear, time-dependent operator $G$.
According to the method of lines the nonlinear, time-dependent operator $G[t,g]$ is discretized
on a fixed grid in phase space,
which turns Eq.~\eqref{eq:GK_symbolic} into a large set of first-order ordinary differential equations
for the time evolution of the state vector $g$.
In this section, explicit RK methods are considered to advance $g_{n}$ at time
$t_{n}$ to $g_{n+1}$ at time $t_{n+1}$ with the timestep $\Delta t=t_{n+1}-t_{n}$.
We focus on explicit RK schemes of the diagonal form
\begin{align}
g_{n+1} & =g_{n}+\Delta t\sum_{j=1}^{s}b_{j}k_{j}\label{eq:RK_scheme}\\
k_{j} & =G\left[t_{n}+a_{j}\Delta t,g_{n}+\Delta t\, a_{j}k_{j-1}\right]\nonumber 
\end{align}
where $s$ is the number of stages and the coefficients fulfill $\sum_{j=1}^{s}b_{j}=1$
as well as $a_{1}=0$. In this simplified scheme, only $k_{j-1}$
is used for computing $k_{j}$, while in general, all $k_{j'}$ with
$j'<j$ can be allowed to contribute. Obviously, this procedure is
memory efficient, since only up to three additional vectors of the
size of $g$ have to be stored. The order of consistency $p$ is
determined by comparing Eq.~\eqref{eq:RK_scheme} with a Taylor expansion
\begin{align}
g_{n+1} & =g_n+\sum_{j=1}^{s}c_{j}\Delta t^{j}\,\frac{\mathrm{d}^j}{\mathrm{d}t^j}g\Big|_{t_{n}}\label{eq:RK_Taylor}\\
 & =g_n+\sum_{j=1}^{p}\frac{\Delta t^{j}}{j!}\,\frac{\mathrm{d}^j}{\mathrm d t^j}g\Big|_{t_{n}}+\sum_{j=p+1}^{s}c_{j}\Delta t^{j}\,\frac{\mathrm d^j}{\mathrm d  t^j}g\Big|_{t_{n}}\nonumber 
\end{align}
of $g$ about $t_{n}$, where the $\{c_{i}\}$ can be computed from
the coefficients $\{a_{j},b_{j}\}$ in Eq.~\eqref{eq:RK_scheme}.
The required order of consistency $p\le s$ thus imposes constraints
on the $\{a_{j},b_{j}\}$. For the linear problem, one inserts the
eigenvalue equation $\mathrm d g/\mathrm dt=\lambda g$ into the Eq.~\eqref{eq:RK_Taylor}
to obtain the stability polynomial 
\begin{align*}
P_{s}(\lambda\Delta t)=1+\sum_{j=1}^{s}c_{j}\,(\lambda\Delta t)^{j}\,,
\end{align*}
allowing to write down the RK stability condition 
\begin{align}
|P_{s}(\lambda\Delta t)|\le1\label{eq:RK_stability_cond}
\end{align}
that must be fulfilled for all $\lambda$ in the left complex half-plane
($\gamma\leq0$) to ensure that these actually stable or damped solutions
are not artificially destabilized by the explicit scheme. Thus, a
sufficiently small timestep must be chosen.
We define $\beta_{\mathrm{imag}}$ ($\beta_{\mathrm{real}}$) to be the extent of the stability boundary
along the imaginary (negative real) axis, i.e. $|P_s(-\beta_{\mathrm{real}})|=|P_s(i\beta_{\mathrm{imag}})|=1$.
For instance, in the case of a simple advection problem $\partial_{t}g=v\,\partial_{x}g$ the
eigenvalues $\lambda=ik_{x}v=i\omega$ are imaginary,
and the maximum timestep is $\Delta t=\beta_{\mathrm{imag}}/\omega_{\mathrm{max}}$.
For maximum time order schemes ($p=s)$, the total stability region
is completely determined by the number of stages. Increasing the number
of stages above $p$ requires additional evaluations of $G[g]$, but 
adds free parameters for shaping the stability polynomial
to lower timestep constraints and reduce the overall computational cost.

Internal stability is found to become
increasingly important at a large number of stages.\citep{HouwenSommeijer1980, Verwer1996}
Since we use diagonal methods with no more than six stages, no restrictions are found in practice.
%In this context it has been stated in Ref.~\citep{Verwer1996} that
%the diagonal method is only useful for less than twelve stages when
%the machine precision is 14 digits. 

Among other choices, GENE features the use of standard $s=p$
second-order (RK2) and fourth-order (RK4) schemes, as well as an optimized
fourth-order scheme (RK4M) with six stages, following Ref.~\citep{Mead1999}.
The corresponding coefficients are given in Table~\ref{tab:RK4_table}
and the stability boundaries are depicted in Fig.~\ref{fig:RK4_RKC_stability}(c).
Additionally, a class of $s$-stage Runge-Kutta-Chebychev
(RKC$s$) schemes is considered, which
are unconditionally unstable for (undamped) waves, but are 
powerful in the case of the real spectrum of the collision operator.
The Chebychev polynomials are defined as
\begin{align*}
T_{0}(x)= & 1\quad T_{1}(x)=x\\
T_{j}(x)= & 2x\, T_{j-1}(x)-T_{j-2}(x),\,\, j\geq2\,.
\end{align*}
Restricting ourselves to first order of consistency $p=1$, shifted Chebychev
polynomials possess the optimal stability along the negative real axis with $\beta_{\mathrm{real}}=2s^2$. 
In order to stabilize small imaginary parts of the $\{\lambda_i\}$,
% which could arise from boundary conditions, for example,
the damped shifted Chebychev polynomials
\begin{align}
P_{{\rm RKC}s}(z)=T_{s}(\omega_{0}+\omega_{1}z)/T_{s}(\omega_{0})\label{eq:RKC_poly}
\end{align}
have been introduced in terms of coefficients $\omega_0=1+\epsilon/s^2$
and $\omega_1=T_{s}(\omega_{0})/T_{s}^{\prime}(\omega_{0})$.\citep{Verwer1996,HouwenSommeijer1980}
Setting the small parameter $\epsilon>0$ then introduces damping. Here, $\epsilon=0.05$ is chosen, which yields
a stability boundary of $\beta_{\mathrm{real}}(s)\approx1.93s^{2}$. 
%, ensures
%that the stability boundary does not intersect the negative real axis
%in the open interval $(-\beta_{\mathrm{real}}(s),0)$, where $\beta_{\mathrm{real}}(s)\approx1.93s^{2}$ is
%the negative real stability boundary of $P_{RKCs}(z)$. 
%In this way, the scheme
%is stable even for small imaginary parts of the eigenvalues $\{\lambda_{i}\}$,
%which could arise from discretization or boundary conditions, for example. 
%
%
\begin{table*}
\begin{centering}
{\footnotesize }%
\begin{tabular}{ccccc}
\hline 
 & {\footnotesize $a_{1}$} & {\footnotesize $a_{2}$} & {\footnotesize $a_{3}$} & {\footnotesize $a_{4}$}\tabularnewline
\hline 
\hline 
{\footnotesize RKC1} & {\footnotesize $0$} & {\footnotesize $1$} &  & \tabularnewline
{\footnotesize RKC2} & {\footnotesize $0$} & {\footnotesize $0.1706953512$} &  & \tabularnewline
{\footnotesize RKC3} & {\footnotesize $0$} & {\footnotesize $0.03569626261$} & {\footnotesize $0.1377742151$} & \tabularnewline
{\footnotesize RKC4} & {\footnotesize $0$} & {\footnotesize $0.03221719644$} & {\footnotesize $-1.987635269$} & {\footnotesize $0.03221719644$}\tabularnewline
\hline 
 & {\footnotesize $b_{1}$} & {\footnotesize $b_{2}$} & {\footnotesize $b_{3}$} & {\footnotesize $b_{4}$}\tabularnewline
\hline 
{\footnotesize RKC1} & {\footnotesize $1$} &  &  & \tabularnewline
{\footnotesize RKC2} & {\footnotesize $0.2497313672$} & {\footnotesize $0.7502686328$} &  & \tabularnewline
{\footnotesize RKC3} & {\footnotesize $0.1115050079$} & {\footnotesize $-0.2891490316$} & {\footnotesize $1.177644024$} & \tabularnewline
{\footnotesize RKC4} & {\footnotesize $0.001610859822$} & {\footnotesize $0.11936272179$} & {\footnotesize $-0.2816076282$} & {\footnotesize $1.160634047$}\tabularnewline
\hline 
\end{tabular}
\par\end{centering}{\footnotesize \par}

\caption{Coefficients for the various (diagonal) first-order Runge-Kutta-Chebychev
methods implemented in the GENE code.\label{tab:RKC_table}}
\end{table*}
The first-order RKC coefficients implemented in GENE are summarized in Table~\ref{tab:RKC_table} and
the stability boundaries are shown in Fig.~\ref{fig:RK4_RKC_stability}(b) and (d).
One recognizes the RKC1 scheme to be identical to the
explicit Euler scheme. The remaining RKC schemes deviate
from the ones described in Ref.~\citep{Verwer1996}.
%, because the latter are non-diagonal
% and require to store two arrays $k_{i-2}$ and $k_{i-1}$ instead of one.
While our (diagonal) approach is more memory
efficient, we lose the opportunity of recursively defining internally
stable schemes for an arbitrary number of stages. However, we observe
in the following sections that at most four stages are necessary in our case.

\section{Timestep optimization with an operator splitting technique\label{sec:Operatorsplit}}

In the previous sections, we have introduced various explicit RK schemes
and discussed properties of the three operators $L[g]$, $N[\bar{\chi},g]$
and $C[g]$ of Eq.~\eqref{eq:GK_symbolic} that determine the stability
of these schemes. In the following, we attempt to find efficient RK
schemes for the individual operators. 
%todo look at this:
In Sec.~\ref{sec:gyrokinetics} it has been shown that for the collisionless part ($N+L$) an extended stability $\beta_{\mathrm{imag}}$ 
along the imaginary axis is required.
If accuracy constraints can be ignored, computational efficiency can be characterized by the ratio $\beta(s)/s$,
as $\beta(s)$ sets the maximum timestep and $s$ measures the cost per step.
Interestingly, among all 4-stage methods the fourth order (RK4) scheme with
$\beta_{\mathrm{imag}}=2\sqrt{2}$ is optimal in that respect.
The (RK4M) $s=6$ scheme has $\beta_{\mathrm{imag}}=4.90$ and thus is about 15\% more efficient than (RK4)
and about 7\% more efficient than the (RK4(3)5[2R+]C) scheme referred to in Ref.~\cite{CandyWaltz2003JCoPhGyro}.
To our knowledge, only a theoretical upper bound of $\beta_{\mathrm{imag}}\leq s$ exists for (even) $s>4$,
and (RK4M) is only 18\% lower than that.\citep{Houwen1977}
%In this respect, the classical fourth-order RK4 scheme is very well
%suited and is superior to e.g. the RK3 method. The optimized fourth-order
%RK4M scheme further increases the efficiency by about $15$\% by using
%six stages. 

The eigenvalues of the collision operator $C[g]$, on the other hand,
extend along the negative real axis. In this case, the RKC methods
discussed in Sec.~\ref{sec:Stability-of-explicitRK} possess a near-optimal
stability polynomial with a computational efficiency $\beta_{\mathrm{real}}/s\approx1.93s$ that increases
linearly in the number of stages.

\begin{figure}
\includegraphics[width=1\columnwidth]{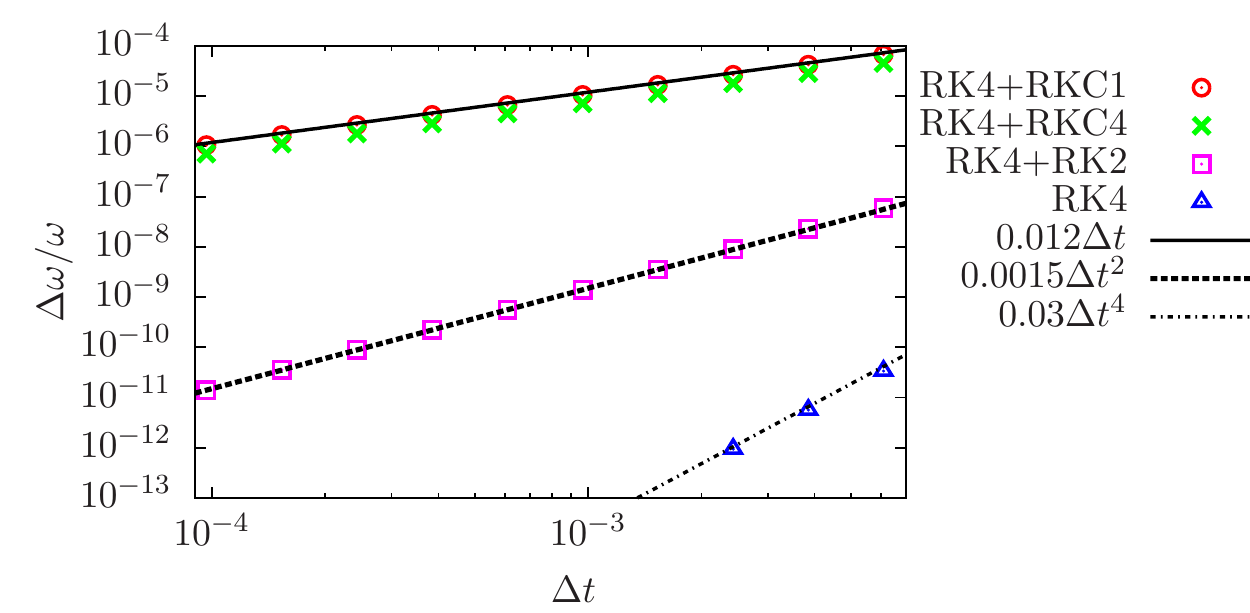}

\caption{The frequency error $\Delta\omega/\omega$ of various explicit RK
schemes is shown as a function of the timestep. The order
of convergence in $\Delta t$ is illustrated by straight lines in
the double logarithmic plot. As expected, the error of the RK4 scheme
(blue triangles) is $O(\Delta t^{4})$. The error of the operator
splitting method Eq.~\eqref{eq:operator_split} is found to be $O(\Delta t^{2})$,
as can be observed when second or higher-order individual schemes are used
(magenta squares). Choosing first-order RKC$s$ schemes (red dots
$s=1$, green crosses $s=4$), the convergence order reduces to $O(\Delta t)$,
but the small prefactor ($\approx10^{-2}$) ensures acceptable accuracy.\label{fig:timeorder}}
\end{figure}

Although it is possible to include the collision operator in $L$ and 
perform time integration with a RK4 method, the
strong benefits of RKC methods can only be exploited when an operator
splitting technique is applied. In exponential notation, it can easily
be shown that the symmetric (Strang) splitting 
\begin{align*}
g_{n+1} & =e^{\Delta t\, C/2}e^{\Delta t\,(N+L)}e^{\Delta\, tC/2}g_{n}\\
 & =e^{\Delta t\,[C+N+L]}g_{n}+O(\Delta t^{3})
\end{align*}
is second-order accurate in $\Delta t$.\citep{Strang1968} In contrast,
the non-symmetric splitting 
\begin{align*}
g_{n+1}=e^{\Delta t\, C}e^{\Delta t\,(N+L)}g_{n}=e^{\Delta t\,[C+N+L]}g_{n}+O(\Delta t^{2})
\end{align*}
is formally only first-order accurate. However, when the propagation
of $g$ with the first operator ($C$) does not change the second
operator ($N+L$), the second half-step can be combined with the first
half-step of the next time iteration. In this case, both of the above
splitting schemes are of second order.\citep{ChengKnorr1976} In gyrokinetics,
this argument holds for linear computations only. In the nonlinear
case, applying collisions on $g$ does generally alter the self-consistent
potentials computed with Maxwell's equations, which in turn changes
the nonlinear operator $N$. 
%A physical interpretation is that the
%potentials have to be kept consistent with the distribution at all times. 
In consequence, second-order accuracy in nonlinear simulations
is expected only for the symmetric splitting. Since the RKC schemes
that we consider here are only first-order accurate anyway, we can
choose the simple approach 
\begin{align}
g_{n+1}^{\mathrm{vl}} & =\mathcal{RK}^{{\rm vl}}\left\{ L[g_{n}]+N[\bar{\chi}_{n},g_{n}],\Delta t\right\} \nonumber \\
g_{n+1} & =g_{n+1}^{\mathrm{vl}}+\mathcal{RK}^{{\rm c}}\left\{ C[g_{n+1}^{\mathrm{vl}}],\Delta t\right\} \label{eq:operator_split}
\end{align}
of alternating propagation with the collisionless (Vlasov) operator
$(L+N)$ and the collision operator $C$, using a common timestep.
The time-stepping schemes $\mathcal{RK}^{{\rm vl}}\{\cdot,\Delta t\}$
and $\mathcal{RK}^{{\rm c}}\{\cdot,\Delta t\}$ can now be chosen
individually. Scanning the timestep for different choices with the
GENE code, Fig.~\ref{fig:timeorder} confirms the above
considerations on a simple test case of an ion temperature gradient
driven (ITG) mode, with kinetic electrons and $\beta_e=0.1\%$.
The frequency error $\Delta\omega=\omega-\omega_{\mathrm{conv.}}$
compared to the converged RK4 result is measured with a precision
of $5\times10^{-13}$ in this example. For an understanding of the
results it is important to note that a $p$th-order scheme has an
$O(\Delta t^{p+1})$ error for computing $g_{n+1}$ from $g_{n}$.
Since the eigenvalue is basically determined by fitting an exponential
as
\[
\lambda+\Delta\lambda=\frac{g_{n+1}-g_{n}+O(\Delta t^{p+1})}{g_{n}\Delta t}\,,
\]
we expect $p$th-order convergence for the frequency error $\Delta\omega$.
As a side note, also the global error for reaching a fixed simulation
time $t$ is $O(\Delta t^{p})$, since choosing a smaller value for $\Delta t$
requires an accordingly larger number of timesteps to be computed.
In summary, using RKC schemes for collisions brings us back to first-order
in time, as expected. However, the observed prefactor of the order
of $10^{-2}$ in Fig.~\ref{fig:timeorder} is relatively small, so
that an acceptable accuracy of at least $10^{-4}$ is obtained.
Thus, in practice, no effect on the accuracy of the physically relevant
solutions is visible with respect to higher-order methods, which has
been confirmed for a large number of linear and nonlinear cases.
%
%%%---------------------figure 2 -------------------------%%%

Two main advantages of applying this operator splitting are identified:
(i) In cases of moderate collisionality, the one stage RKC1 scheme
is sufficient for collisions, which saves three(five) calls of $C[g]$,
compared to including it in the RK4(RK4M) scheme. This remains valid
in nonlinear computations, although the total fraction of CPU time
spent to compute collisions will be smaller.
(ii) In strongly collisional
cases, the timestep is restricted by strongly damped collisional eigenvalues.
Computation time can then be saved by adding RKC stages, which allows
larger timesteps. In nonlinear simulations the timestep is rarely
dominated by collisions, even if it is in the corresponding linear case.

Thus, it is reasonable to determine the optimal number of RKC stages
dynamically, so that the maximum stable RKC timestep $\Delta t_{\mathrm{RKC}}$
is always somewhat larger than $\Delta t_{\mathrm{RK4}}$, the maximum
stable RK4 timestep. In this way, collisions never restrict the timestep.
This makes sense, as long as evaluating the collisionless part dominates
the computational cost, which is generally the case.
For evaluating $\Delta t_{RK4}$ and $\Delta t_{RKC}$, a small set of most restrictive eigenvalues
of $L$ and $C$, $\lambda_L$ and $\lambda_C$ are pre-computed with fast largest-magnitude SLEPc algorithms.
When multiple Fourier modes $k_y$ are present, we make use of the fact that $L$ and $C$ are block-diagonal in this dimension.
In this way, only a very low percentage of the following initial value computation ($\simleq1\%$ of a linear run)
is needed for this step.
%Note that the computation of physically interesting largest real eigenvalues is not as fast,
%which can be attributed to the shape of the spectrum.
%In general nonlinear simulations, the timestep $\Delta t$
%is not constant, but rather develops in time. Thus, it proves useful
%to dynamically set the number of RKC stages to a minimum, which still
%ensures that the timestep is never restricted by collisions.

We note, however, that the equations implemented in GENE
require the computation of $f$ (and $\bar{\chi}$) before every call
of $C$, because these fields have to be kept consistent with $g$.
This produces an additional overhead of the splitting scheme.
In rare cases, the computation of $f$ and $\bar{\chi}$ is found to
be relatively costly, but in general this is easily over-compensated
by the gain in timestep or the less frequent calls of $C$ itself.
A positive side-effect is related to the fact
that in the GENE code the velocity space dimension $\mu$ is stored
in the last index of $g$ and therefore is widely spread in
the system memory. Since only collisions require ghost-cells in this
dimension, which are exchanged via the message passing interface library,
the reduction in number of calls of $C$ improve the parallelization efficiency. 

\begin{table*}
\centering{}\emph{\small }%
\begin{tabular}{cccc}
\hline 
{\small MTM} & {\small $\Delta t\times100(R/c_{s})$} & {\small time$/s$} & {\small speedup}\tabularnewline
\hline 
{\small RK4} & {\small $0.107$} & {\small $1181$}    & {\small 1}   \tabularnewline
{\small +RKC1} & {\small $0.0772$} & {\small $1099$} & {\small 1.1} \tabularnewline
{\small +RKC2} & {\small $0.246$} & {\small $440$}   & {\small 2.7} \tabularnewline
{\small +RKC3} & {\small $0.246$} & {\small $529$}   & {\small 2.2} \tabularnewline
{\small +RKC4} & {\small $0.246$} & {\small $620$}   & {\small 1.9} \tabularnewline
\hline 
{\small RK4M} & {\small $0.124$} & {\small $1534$}   & {\small 0.8} \tabularnewline
{\small +RKC1} & {\small $0.0772$} & {\small $1490$} & {\small 0.8} \tabularnewline
{\small +RKC2} & {\small $0.301$} & {\small $459$}   & {\small 2.6} \tabularnewline
{\small +RKC3} & {\small $0.426$} & {\small $376$}   & {\small 3.1} \tabularnewline
{\small +RKC4} & {\small $0.426$} & {\small $432$}   & {\small 2.7} \tabularnewline
\hline 
\end{tabular}\emph{\small \hspace{0.1\textwidth}}%
\begin{tabular}{cccc}
\hline 
{\small TEM} & {\small $\Delta t\times100(R/c_{s})$} & {\small time$/s$} & {\small speedup}\tabularnewline
\hline 
{\small RK4} & {\small $0.149$} & {\small $246$}   & {\small 1 }\tabularnewline
{\small +RKC1} & {\small $0.148$} & {\small $152$} & {\small 1.6}\tabularnewline
{\small +RKC2} & {\small $0.148$} & {\small $207$} & {\small 1.2}\tabularnewline
{\small +RKC3} & {\small $0.148$} & {\small $266$} & {\small 0.9}\tabularnewline
{\small +RKC4} & {\small $0.148$} & {\small $309$} & {\small 0.8}\tabularnewline
\hline 
{\small RK4M} & {\small $0.257$} & {\small $212$}  & {\small 1.2}\tabularnewline
{\small +RKC1} & {\small $0.257$} & {\small $113$} & {\small 2.2}\tabularnewline
{\small +RKC2} & {\small $0.257$} & {\small $146$} & {\small 1.7}\tabularnewline
{\small +RKC3} & {\small $0.257$} & {\small $179$} & {\small 1.4}\tabularnewline
{\small +RKC4} & {\small $0.257$} & {\small $212$} & {\small 1.2}\tabularnewline
\hline 
\end{tabular}{\caption{\label{tab:RKCa_lin}\emph{Test cases for the use of operator splitting with the adaptive RKC
method combined with the RK4 and RK4M schemes.
Compared to the RK4 case without splitting, the code efficiency is enhanced by up
to a factor of three, depending on the problem. Simulation times are
determined on 64 CPUs of the HELIOS system.}}%\citep{heliosweb}
}
\end{table*}
The efficiency of the adaptive RKC operator splitting methods in combination
with RK4 and RK4M is demonstrated in the following. We first
focus on linear physics and compute the fastest growing solution with the initial value solver.
Two typical cases that are sensitive to collisions are chosen.
One is a trapped electron mode problem (TEM) in circular model geometry.
The other is a microtearing mode problem (MTM) for physics parameters of the
{ASDEX} Upgrade discharge 27963 at the radial position $\rho_{\mathrm{tor}}=0.85$.
The eigenvalues computed by the GENE code are 
$\gamma_{{\rm TEM}}=0.2610\, c_s/L_{\mathrm{ref}}$, $\omega_{{\rm TEM}}=-0.6380\, c_s/L_{\mathrm{ref}}$ for the TEM case,
and $\gamma_{{\rm MTM}}=0.181\, c_s/L_{\mathrm{ref}}$, $\omega_{{\rm MTM}}=-1.332\, c_s/L_{\mathrm{ref}}$ for the MTM case. 
They coincide for all schemes considered up to the given convergence accuracy of $10^{-3}$.
%although the splitting scheme is only first-order accurate in time.
Table~\ref{tab:RKCa_lin} summarizes the results for code efficiency. In both cases, operator
splitting leads to a strong increase in efficiency. 
In the TEM case with $\nu_{ei}/\omega=0.38$,
the timestep is not limited by collisions. Here, less frequent calls
of the collision operator lead to shorter runtime and thus RKC1 is most efficient.
In the MTM case, the collisionality is larger ($\nu_{ei}/\omega=1.2$), so that adding up to two
or three RKC stages increases efficiency due to a gain in timestep.
Interestingly, the RKC1 splitting method requires a smaller timestep
(which is explained by the lower stability boundary) but still is
slightly more efficient due to the reduced number of calls of the
collision operator. 
As expected, for both cases the runtime increases
as soon as the optimal number of RKC stages is exceeded.

The benefits of operator splitting with RKC schemes become even more striking when
replacing the RK4 method with the optimized six-stage RK4M method
that is more efficient for large imaginary eigenvalues. Again, this
is attributed to the use of a larger timestep, which also results
in less overall evaluations of the collision operator. In consequence,
the combination of RK4M and RKC1 leads to the lowest runtime for these
linear runs. Compared to the standard RK4 scheme, the speedup is a
factor of two to three, depending on the parameter set. This is
significant when comprehensive physics models are employed to perform large (multidimensional) parameter studies,
as is routinely done in quasilinear transport
predictions for fusion plasmas (see \citep{Dannert05PoP,Bourdelle2007,Merz2010,Casson2013},
for example).
 
Additionally, we have compared the time traces of a nonlinear simulation
of the TEM case with and without operator splitting using up to two
stages. The simulation times are given in Table~\ref{tab:RKCa_nonlin}.
Here, the timestep is set by linear physics of the high wavenumbers
even in the nonlinear simulation. Due to reduced computational time
per step, the RKC1 operator splitting method in combination with the
RK4M time scheme has the largest speedup with respect to the RK4 scheme
without operator splitting. We note that the physical results are
identical within the statistical error bars inherent to nonlinear
turbulence simulations. For the present MTM case the nonlinear terms
dominate the timestep limit, as shown in the next Section, similarly
to previously published MTM simulations in a slightly different parameter
regime.\citep{Doerk11} In such cases, the RKC1 collision scheme is
most efficient. 
\begin{table}
\emph{\small }%
\begin{tabular}{cccc}
 & {\small $\langle \Delta t\rangle \times10^{3}(R/c_{s})$} & {\small CPU time/step$[s]$} & {\small speedup}\tabularnewline
\hline 
\hline 
{\small RK4} & {\small 1.74 } & {\small 1.08 } & {\small 1 }\tabularnewline
{\small +RKC1} & {\small 1.74 } & {\small 0.74 } & {\small 1.45}\tabularnewline
{\small +RKC2} & {\small 1.74 } & {\small 0.96 } & {\small 1.12}\tabularnewline
\hline 
{\small RK4M} & {\small 3.01 } & {\small 1.59 } & {\small 1.17}\tabularnewline
{\small +RKC1} & {\small 3.01 } & {\small 0.99 } & {\small 1.87}\tabularnewline
{\small +RKC2} & {\small 3.01 } & {\small 1.23 } & {\small 1.51 }\tabularnewline
\hline 
\end{tabular}\emph{\caption{Nonlinear simulation times for the TEM test
case, measured on 512 CPUs of the HELIOS system.
%\emph{\citep{heliosweb}}
Since the timestep is dynamically adapted, $\langle \Delta t\rangle$ is time-averaged.
\label{tab:RKCa_nonlin}}
}
\end{table}

\section{Timestep optimization in nonlinear simulations\label{sec:nlev}}
%
%-----------------figure 3 -----------------%%
\begin{figure}
\begin{centering}
\includegraphics[width=0.8\columnwidth]{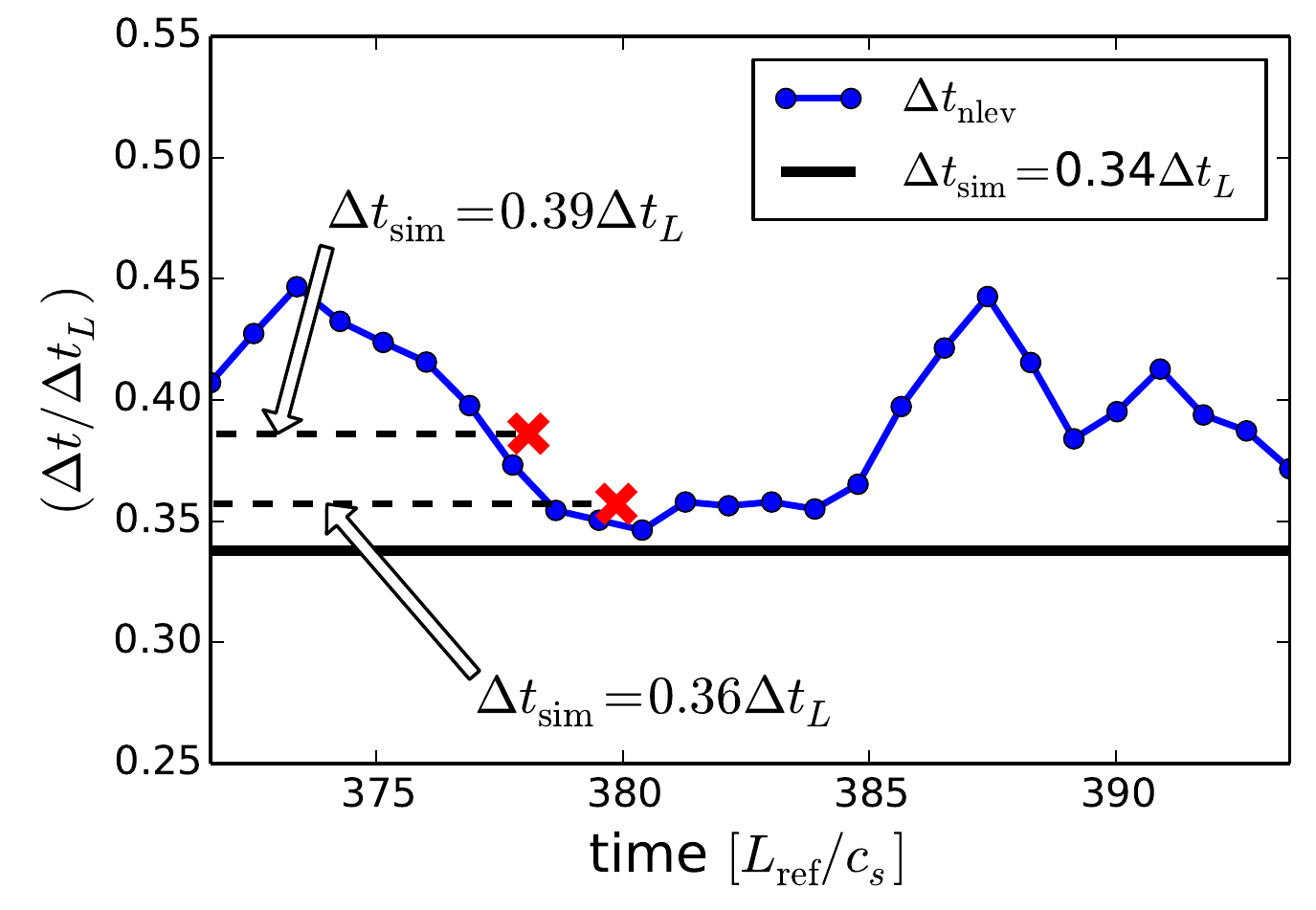}

\caption{
This figure demonstrates that the eigenvalues with linearized $v_{\chi}$
terms are meaningful for stability of the RK4 scheme.
As an example, an electromagnetic ITG/TEM simulation in {ASDEX} Upgrade outer core geometry is chosen. 
The timesteps $\Delta t_L$ and $\Delta t_{\mathrm{nlev}}$ are computed by
inserting $\lambda_L$ and $\lambda_{\mathrm{nlev}}$ into the stability
condition Eq.~\eqref{eq:RK_stability_cond}, respectively. The linear
timestep is $\Delta t_{L}=0.00518\, L_{\mathrm{ref}}/c_s$ in this case. While the simulation
is stable when the timestep is just below $\Delta t_{\mathrm{nlev}}$,
a numerical instability is detected in two other simulations (marked
by dashed lines ending in red crosses),
when $\Delta t_{\mathrm{sim}}$ exceeds $\Delta t_{\mathrm{nlev}}$.\label{fig:nlev_dt_stability}}
\par\end{centering}
\end{figure}

This section addresses the modification of the maximum stable timestep by the nonlinear term Eq.~\eqref{eq:nonlinearity}.
For simplicity, in the present section we will
include collisions in the linear operator $L$, unless $C$ is 
separated from the RK4 scheme, as described in the previous section.
In this sense, we denote $\Delta t_{L}$ the timestep obtained by considering
only the linear terms. The nonlinear term will be analysed in a linearized
form $N[\bar{\chi}_{0},g]$, which is obtained by freezing the potential
to a snapshot of $\bar{\chi}_{0}=\bar{\chi}(t_{n})$ taken at the
current time $t_{n}$. This procedure allows to access the largest
magnitude eigenvalues $\lambda_{\mathrm{nlev}}$ of the combined operator
$L[g]+N[\bar{\chi}_{0},g]$, from which the maximum timestep $\Delta t_{\mathrm{nlev}}$
can be computed exactly from Eq.~\eqref{eq:RK_stability_cond}. Since the drift velocities $v_{\chi}^{x}$
and $v_{\chi}^{y}$ do not change much during one step, $\Delta t_{\mathrm{nlev}}$
is expected to accurately describe the stability limit of the RK scheme
in the nonlinear regime. We test our hypothesis by performing three
simulations with fixed timestep and identical initial condition.
As illustrated in Fig.~\ref{fig:nlev_dt_stability}, the scheme becomes unstable,
when $\Delta t_{\mathrm{nlev}}$ sinks below the
simulation timestep. Up to this point the time-traces of physical
quantities are identical, reflecting the fact that the results are
converged with respect to the timestep. 
%
%
%--------------figure 4--------------_%%
\begin{figure*}
\begin{centering}
(a)\hspace{0.33\textwidth}(b)\hspace{0.33\textwidth}(c)
\par\end{centering}

\includegraphics[width=0.27\textwidth]{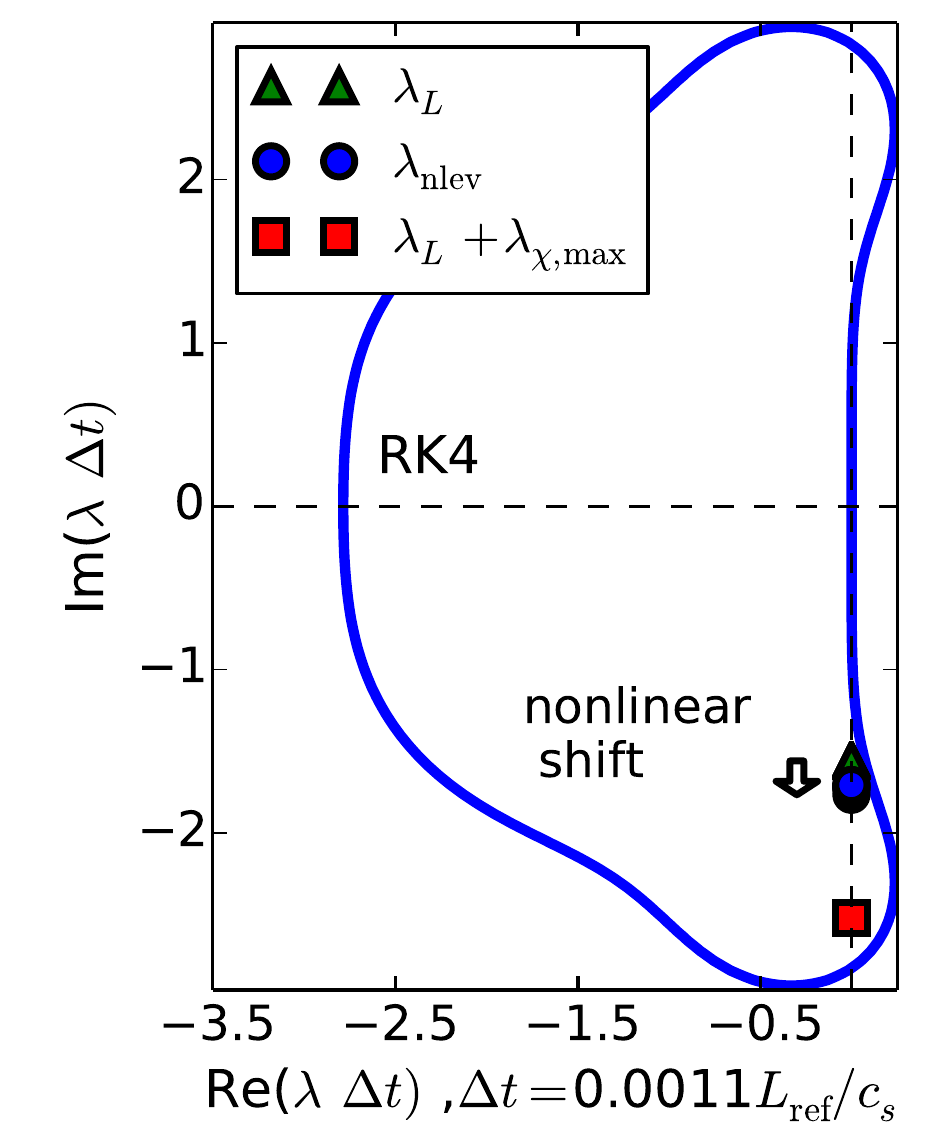}\hfill{}\includegraphics[width=0.27\textwidth]{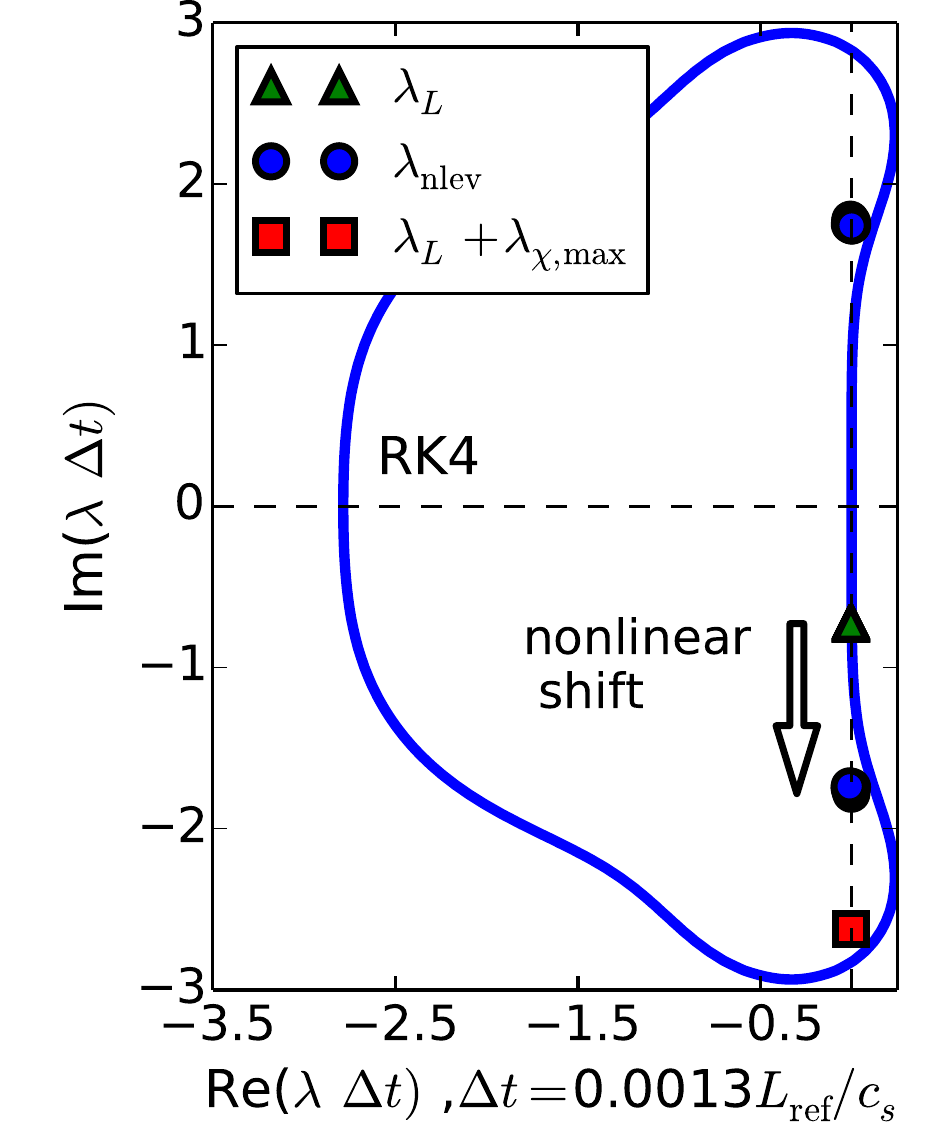}\hfill{}\includegraphics[width=0.27\textwidth]{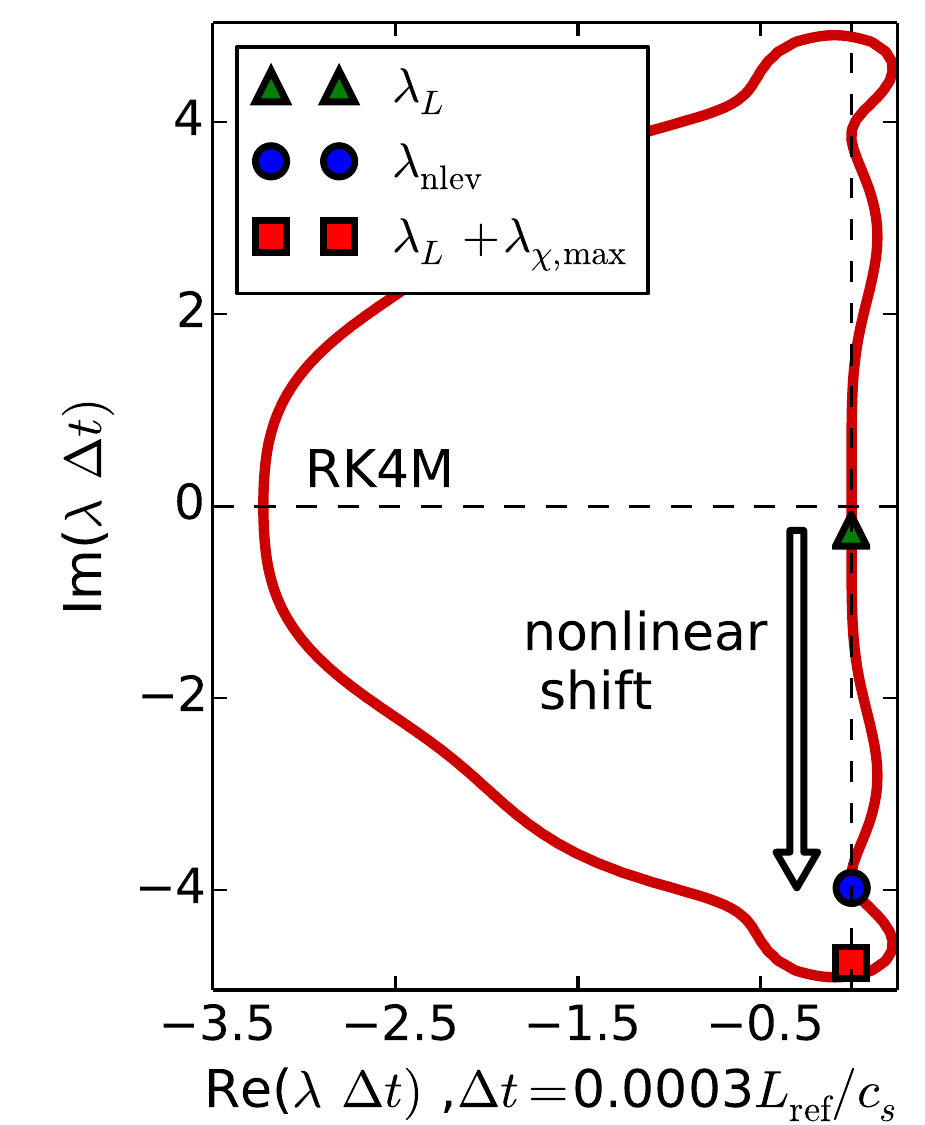}

\caption{This figure illustrates the position of the most timestep-relevant eigenvalues $\lambda_{L}$ and $\lambda_{{\rm nlev}}$
at one point in time. The sum of $\lambda_{L}$ and the maximum nonlinear shift $\lambda_{\chi,{\rm max}}$ is also depicted,
accounting for the sign of $\lambda_{L}$.
Three parameter sets with increasing nonlinear contribution are chosen.
(a) is an ITG/TEM case in circular geometry, (b) and (c) use realistic tokamak geometry of the
{ASDEX} Upgrade discharges \#26459 and \#27963, respectively. 
The shift of $\lambda_{{\rm nlev}}$ with respect to $\lambda_{L}$ is along the imaginary axis, as expected.
%Note that the CFL-type timestep $\Delta t_{\chi,\mathrm{max}}=\beta_{\mathrm{imag}}/\lambda_{\chi,\mathrm{max}}$ in Eqs.~\eqref{eq:taucomb} and \eqref{eq:tau_cfl} is based on the (estimated) shift from green triangles to red squares.
Note that Eq.~\eqref{eq:tau_cfl} uses an effective eigenvalue $\lambda_{\chi,\mathrm{max}}/c_{\mathrm{cfl}}$. Setting $c_{\mathrm{cfl}}=0.3$ to a typical value, this lies (far) outside the figure in cases (b) and (c), which demonstrates the improvement of Eq.~\eqref{eq:taucomb}.
\label{fig:nlev_model}}
%nlev/bigtest3
%nlev/AUG26459
\end{figure*}
%

%Thus, it seems that we have a powerful tool at hand to allow an exact
%computation of the maximum stable timestep.% in nonlinear simulations.
Thus, it seems that we have a powerful tool at hand to exactly compute the
maximum stable timestep, even in nonlinear simulations.
However, performing this kind of ``nonlinear eigenvalue computation''
is not feasible at every timestep. In the following, we develop a
fast method for approximating the influence of the nonlinear term on the eigenvalue spectrum.
The nonlinearity itself constitutes a pure advection problem.
We thus expect the eigenvalues to spread along the imaginary axis.
Also the timestep restrictions for the linear operator stem from
eigenvalues with a dominating imaginary part. Simply adding the eigenvalues
is tempting, but of course the eigenvalues of the sum of two
operators can not be obtained as the sum of the eigenvalues of the
two separate operators, unless they are diagonalized by the same unitary
transform.

%Nevertheless, in practice, a quick estimate of the nonlinear eigenvalues
%and a method of combining them with the eigenvalues of the linear
%operator are needed. 
Nevertheless, for combining linear and nonlinear effects, we replace Eq.~\eqref{eq:nonlinearity} by
\begin{align}
N[\bar{\chi},g]&\rightarrow N_{{\rm max}}[g]=i\lambda_{\chi,{\rm max}}g,\nonumber\\
\lambda_{\chi,{\rm max}}&=|v_{\chi,{\rm max}}^{x}|k_{x,{\rm max}}+|v_{\chi,{\rm max}}^{y}|k_{y,{\rm max}}
\label{eq:lambdachimax}
\end{align}
where the real number $\lambda_{\chi,{\rm max}}$
maximizes the advection velocity over $\{x,y,z,\mu\}$ phase space
as well as species and uses the largest wave-vectors $k_{x}$ and
$k_{y}$ present in the simulation.
Fortunately, $\lambda_{\chi,{\rm max}}$ can be computed at every timestep with
negligible effort. Because we are now dealing with a constant advection,
the resulting total equation $\partial_{t}g=L[g]+N_{{\rm max}}[g]$
indeed is solved by the Ansatz $g(t)=g_{0}\exp[i\lambda t]$ with
$\lambda=\lambda_{L}+\lambda_{\chi,{\rm max}}$. The combined timestep
$\tilde{\Delta t}_{\mathrm{comb}}$ is defined by finding the roots
of $|P_{s}[(\lambda_{L,{\rm max}}+\lambda_{\chi,{\rm max}})\tilde{\Delta t}_{\mathrm{comb}}]|=1$
with $\lambda_{L,{\rm max}}$ being a set of most restrictive eigenvalues
of $L$ that are pre-computed before the time-stepping starts. Defining
$\Delta t_{\chi,{\rm max}}=\beta_{\mathrm{imag}}/\lambda_{\chi,{\rm max}}$, the
combined timestep is also well captured by setting 
\begin{align}
\Delta t_{{\rm comb}}=1/(\Delta t_{L}^{-1}+\Delta t_{\chi,{\rm max}}^{-1})\,.\label{eq:taucomb}
\end{align}
%Although here the (small) real part of the linear eigenvalues is ignored,
%we find $\tilde{\Delta t}_{\mathrm{comb}}\approx\Delta t_{\mathrm{comb}}$
%and thus we use the latter, for simplicity.
For completeness we note that, in the GENE code, too frequent timestep changes are
avoided by using a threshold of about 5 per cent. For $\Delta t$
to be adapted, the current estimate must deviate from the present
value by more than this threshold. Additionally, successive timestep increments
are only allowed after a minimum of 200 steps.

The presented scheme is indeed robust under the following three conditions.
(i) $\lambda_L$ is dominantly imaginary.
(ii) The shift of $\lambda_L$ is along the imaginary axis.
(iii) $\lambda_{\chi,\mathrm{max}}$ overestimates the nonlinear shift by more than the above mentioned threshold value.
The first condition (i) is justified by the use of non-dissipative differencing (plus a negligible real part caused by hyperdiffusion).
%added section about Merz2008
On the second condition (ii) we note that
the time-average of $N$ can be modelled as a diffusive term
that balances linear growth.\citep{Merz2008}
This would intuitively imply an eigenvalue shift along the negative real axis.
Indeed, nonlinear $E_{\chi}\times B$ advection mediates dissipation by transporting 
fluctuation energy from driven phase space regions into dissipative regions.
While this mechanism is important for the formation of a statistically stationary turbulent state,
$N$ is purely advective at each point in time, which is relevant for the stability of time integration.
%end of added section
The third condition (iii) holds, because 
$\lambda_{\chi,\mathrm{max}}$ is a global maximum over phase space and thus overestimates the actual stability
restriction of the nonlinear term. Also the interplay with linear
terms has been observed to lower the stability constraint in some cases. However, the theoretical maximum
can closely be reached (cases with dominant $A_{1\|}$ fluctuations and fine $k_x$ resolution show this behavior).
Nevertheless, we allow $\Delta t_{\chi,\mathrm{max}}$ to be multiplied with a user specified constant
$c_{\mathrm{cfl}}$, which can be set larger than unity in most cases.
Robustness for the general case is obtained for $c_{\mathrm{cfl}}=0.95$ to
compensate the threshold mentioned above. 
It should be noted that a non-spectral treatment of the $x$ ($y$) dimension demands a correction factor to $k_x$ ($k_y$) in Eq.~\eqref{eq:lambdachimax},
which can be deduced from the corresponding differencing scheme.
%
%%%---------------------figure 5---------------------------%%%
\begin{figure*}
\begin{centering}
(a)\hspace{0.33\textwidth}(b)\hspace{0.33\textwidth}(c)
\par\end{centering}

\includegraphics[width=0.3\textwidth]{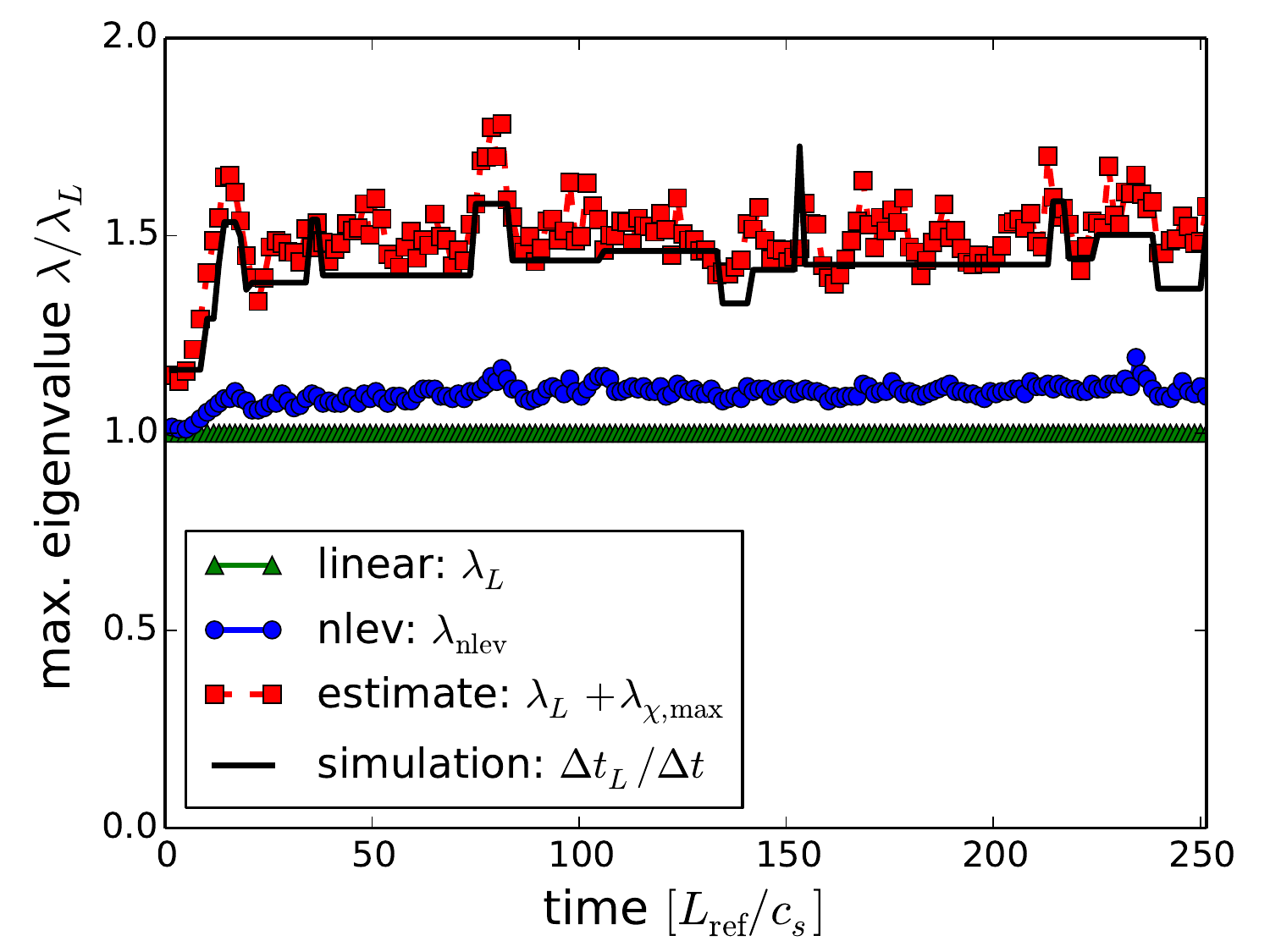}\hfill{}\includegraphics[width=0.3\textwidth]{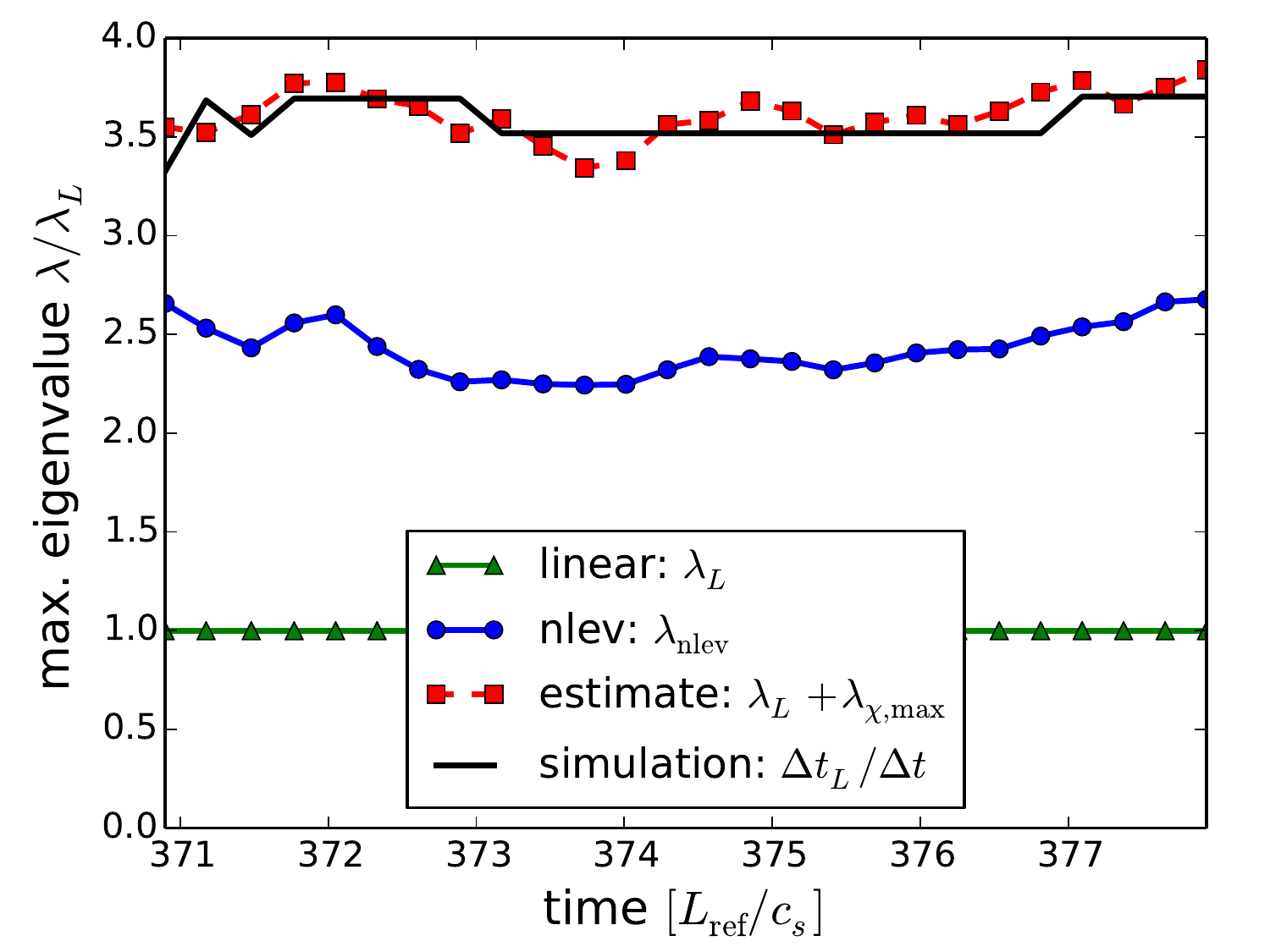}\hfill{}\includegraphics[width=0.3\textwidth]{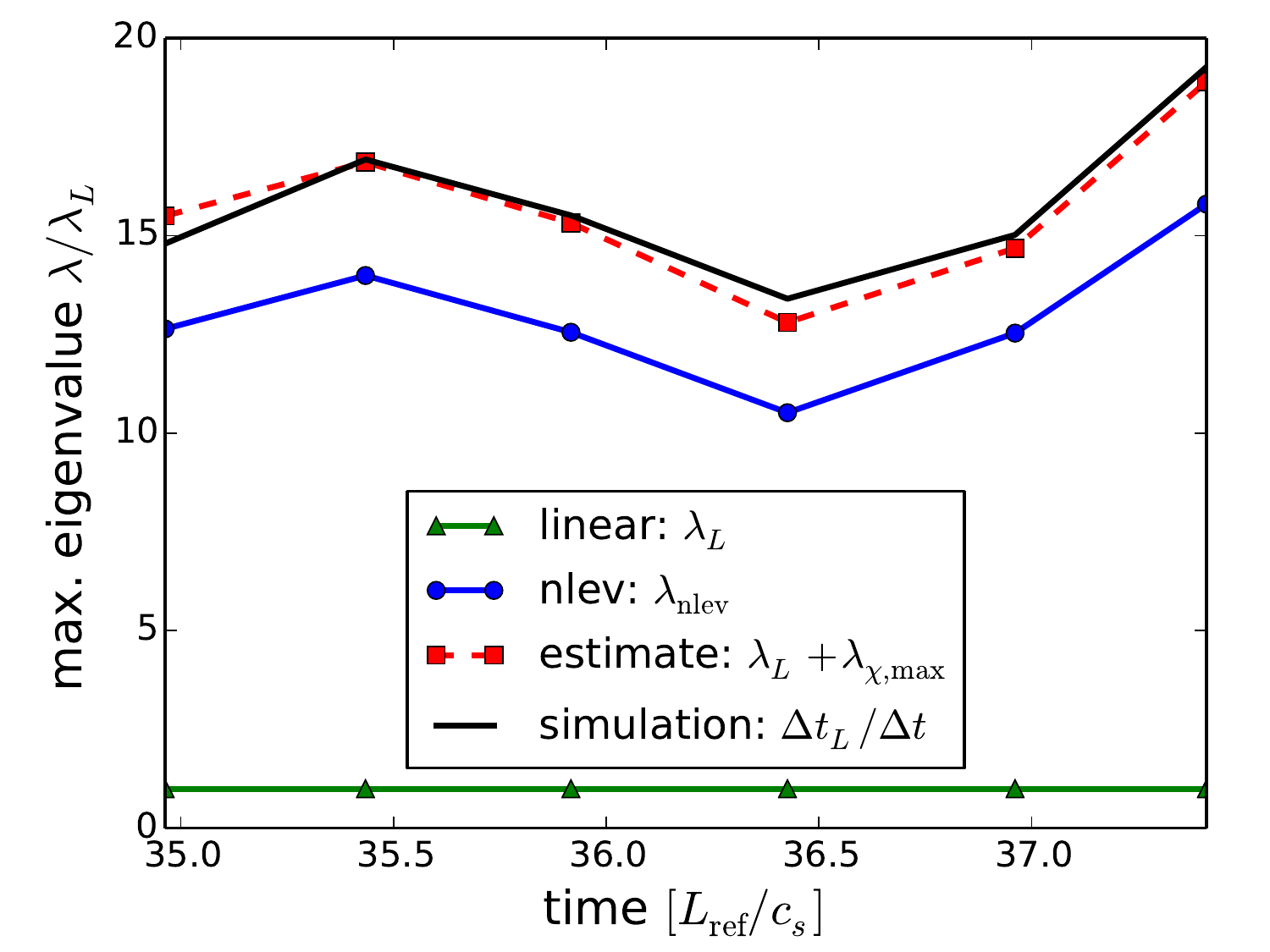}

\caption{Time traces of linear, estimated nonlinear, and exact nonlinear eigenvalues
for the three sets of parameters of Fig.~\ref{fig:nlev_model} The
sum $\lambda_{L}+\lambda_{\chi,{\rm max}}$ overestimates $\lambda_{{\rm nlev}}$ at any time,
leading to a robust timestep limit. The corresponding time trace
of $1/\Delta t$ does not exactly match the estimate due to built-in
thresholds for timestep changes, but it stays quite close.
%Note that a timestep based on $\lambda_{\chi,\mathrm{max}}$ alone (Eq.~\eqref{eq:tau_cfl}) has little to do with the real situation.
Here, the $\lambda_{{\rm nlev}}$ are only computed each 1000th timestep.
\label{fig:nlev_timetrace}}
\end{figure*}

The shift of the eigenvalues due to the nonlinearity as well as our model
are illustrated in Fig.~\ref{fig:nlev_model} for three distinct
cases with increasing nonlinear contribution. The linear eigenvalues
are shifted along the imaginary axis, as expected from the advective character of $N$.
Using the same three parameter sets, we show in Fig.~\ref{fig:nlev_timetrace} that
the sum of the maximum linear eigenvalue and the maximum nonlinear
shift $\lambda_{L}+\lambda_{\chi,{\rm max}}$ is indeed always overestimating
the exact result $\lambda_{{\rm nlev}}$. 
In the third case, the exact result is only overestimated by about 20\% and
also other cases showed almost no difference between modeled and real timestep limits. 
%Overall, there seems to be still some room for improvement in the first two examples. 
As mentioned above, other cases allow a $c_{\mathrm{cfl}}$ prefactor to 
$\Delta t_{\chi,{\rm max}}$ to be set larger than one.
While this increases the efficiency, general robustness is lost.
%One may introduce a user-specified reduction
%factor for $\lambda_{\chi,{\rm max}}$ to adjust the adaptation scheme
%to a specific simulation. Such a measure destroys general robustness
%but allows to save computation time if stability should still be maintained.
A less strict, yet robust automatic adaptation scheme would of course
be desirable, but this seems to be impossible to find without performing
actual expensive eigenvalue computations.

However, we want to point out that the presented estimate of Eq.~\eqref{eq:taucomb}
is already superior to the previously used CFL method 
\begin{align}
\Delta t_{\mathrm{cfl}}=\text{min}(\Delta t_{L},c_{\mathrm{cfl}}\times\Delta t_{\chi,{\rm max}})\,.\label{eq:tau_cfl}
\end{align}
that uses the timestep $\Delta t_{\chi,{\rm max}}$ multiplied with
a CFL constant $c_{\mathrm{cfl}}\sim0.3-0.5$ and never exceeds the linear timestep.
Importantly, Eq.\eqref{eq:tau_cfl} requires $c_{\mathrm{cfl}}$ to be set smaller than one
in order to stabilize cases of weak (but non-negligible) nonlinear influence.
%Having the nonlinearly shifted eigenvalues in mind.
%mind, it is obvious that in cases of weak (but non-negligible) nonlinear
%influence such a CFL safety factor is not sufficient and can lead
%to instability.
In the opposite limit of dominating nonlinear dynamics,
Eq.~\eqref{eq:tau_cfl} produces unnecessarily small timesteps. Combining
the maximum eigenvalues according to Eq.~\eqref{eq:taucomb}, naturally
captures both limits in a satisfactory way.

Coming back to our exemplary cases of Figs.~\ref{fig:nlev_model}
and \ref{fig:nlev_timetrace}, we observe that in case (a), which
is am ITG/TEM case in circular geometry, the nonlinear term has only
a weak influence and the improved estimate yields about the same timestep
as Eq.~\eqref{eq:tau_cfl} with $c_{\mathrm{cfl}}=0.3$ would provide.
However, the latter becomes unstable for larger $c_{\mathrm{cfl}}$
factors, suggesting to set $c_{{\rm cfl}}=0.3$ by default. In the
second example (b), which is a realistic ITG/TEM setup, $\Delta t_{\mathrm{comb}}$
is about a factor or two larger than $\Delta t_{\mathrm{cfl}}$ (with
$c_{\mathrm{cfl}}=0.3$). Example
(c) is a realistic ITG/MTM mixed case, in which the nonlinear timestep
limit is even stronger, so that the linear terms play almost no role. In
such cases, any $c_{\mathrm{cfl}}<1$ will reduce the code efficiency
by about the same factor.
In future high-temperature devices like ITER, the normalized fluctuation amplitude ($v_{\chi}$) is expected to be smaller.
However, the impact on the timestep is also determined by resolution settings.
Beyond the examples given here, many other realistic gyrokinetic simulations
show a significant nonlinear timestep restriction.
Thus, our improved estimate can save a substantial amount (up to two thirds) of computation time.

%todo make remark here: CBC has less nonliear effect. Including realistic tokamak geometry and 
% using high resolution in $k_\perp$ space
%As a final remark, we note that gyrokinetic simulations that require a fine perpendicular grid,
%which is often given in in realistic geometry, when substantial magnetic field fluctuations are involved,
%or in multiscale studies, tend to show a stronger nonlinear contribution to the timestep.
%As a final remark, we note that the timestep in many realistic setups is observed 
%to be strongly influenced by nonlinear dynamics.
%At the same time, one is usually not willing to experiment with increasing a $c_{\mathrm{cfl}}$
%factor above one and risk instability because one single turbulence
%simulation can consume several hundreds or even millions of CPU-hours
%on present-day supercomputers. 
%Thus, our improved estimate can save a substantial amount (up to two thirds) of computation time.

\section{Conclusions}

In summary, we presented two methods for increasing the efficiency of
gyrokinetic simulations and applied these to the plasma turbulence code GENE.
First, we matched individual explicit Runge-Kutta schemes to the properties
of individual parts of the equation by applying an operator splitting
technique. For the collisionless part we chose classical and advanced
fourth-order schemes. For collisions, we restricted ourselves to first-order
Runge-Kutta-Chebychev schemes, since they possess optimal stability
properties, while higher-order schemes offer much smaller efficiency
gains. Thereby, we reached an increased timestep and/or fewer evaluations
of the collision operator, resulting in a speedup by a factor of up
to three both in strongly and weakly collisional cases. A possible
application is given by extremely large (multi\-dimensional) parameter
studies, which are, for example, needed in quasilinear transport modeling
of tokamak plasmas. Time savings due to our method are striking especially
in the tokamak edge, where the collisionality is increased.

Second, we investigated the impact of nonlinear advection on the timestep.
Based on the observation of a frequency-shift in the eigenvalue spectrum due to the
$E_{\chi}\times B$ advection velocity (which is an interesting topic
in itself), we developed an improved and robust timestep estimate
for nonlinear simulations. 
Beyond the examples shown in this paper, the new adaptation scheme has been successfully
applied to a large number of simulations.
Avoiding unnecessarily small timesteps,
a speedup of up to a factor of two to three is realized for realistic
problems. 
This is particularly important for large simulations including
comprehensive physics, experimental plasma shaping, kinetic electrons,
multiple scales, and possibly also profile variations. Constituting
the high-end of fusion plasma modeling, such simulations yield the
most accurate description of plasma turbulence currently
available, but they are expensive: one run can consume millions of
CPU hours on present-day supercomputers.

Since the choice of algorithms and their implementation are already
highly optimized in GENE (as in other state-of-the-art
codes), this further increase of efficiency is really significant.
The techniques discussed in this work can prove extremely useful,
also for other simulation codes with similar numerical schemes.

\section*{Acknowledgments}

The authors would like to thank J.~Abiteboul, T.~Dannert, T.~G\"orler,
D.~R.~Hatch, F.~Merz, E.~Sonnendr\"ucker, and D.~Told for fruitful
discussions and M. Dunne for extracting code input from the {ASDEX}
Upgrade database.
This work was supported by the Nu-FuSE project which
is funded through G8 Multilateral Research by Funding Nu-FuSE grant
JE 520/4-1.
The research leading to these results has received funding from the 
European Research Council under the European Union's Seventh Framework
Programme (FP7/2007-2013)/ERC Grant Agreement No. 277870.
The numerical results presented in this work were carried
out using the HELIOS supercomputer system at the Computational Simulation
Centre of International Fusion Energy Research Centre (IFERC-CSC),
Aomori, Japan, under the Broader Approach collaboration between Euratom
and Japan, implemented by Fusion for Energy and JAEA and using the
resources of the RZG computing center, Garching, Germany. 

%\section*{References}

%\bibliographystyle{elsarticle-num}
%\bibliographystyle{model3-num-names}
%\bibliographystyle{model1a-num-names}
%\bibliography{bibliography_nlev}

\end{document}